\documentclass{fic-l}

\def\ov#1{\overline{#1}}

\def\wt#1{\widetilde{#1}}
\def\vb#1{\mbox{\boldmath$#1$}}
\def\pd#1#2{\frac{\partial #1}{\partial #2}}
\def\fd#1#2{\frac{\delta #1}{\delta #2}}
\def\wh#1{\widehat{#1}}
\def\bdot{\,\vb{\cdot}\,}
\def\btimes{\,\vb{\times}\,}

\def\bhat{\wh{{\sf b}}}
\def\cal#1{\mathcal{#1}}
\def\bf#1{\mathbf{#1}}

\newcommand{\bc}{\begin{center}}
\newcommand{\ec}{\end{center}}
\newcommand{\bt}{\begin{tabbing}}
\newcommand{\et}{\end{tabbing}} 
\newcommand{\be}{\begin{eqnarray*}}
\newcommand{\ee}{\end{eqnarray*}}
\newcommand{\bs}{\begin{slide}}
\newcommand{\es}{\end{slide}}

\begin{document}

\title{Variational Formulations of Exact and Reduced Vlasov-Maxwell Equations}

\author{Alain J.~Brizard}
\address{Department of Physics, Saint Michael's College \\ One Winooski Park, Colchester, Vermont 05439, USA}
\email{abrizard@smcvt.edu}



\begin{abstract}
The foundations of gyrokinetic theory are reviewed with an emphasis on the applications of Lagrangian and Hamiltonian methods used in the derivation of
nonlinear gyrokinetic Vlasov-Maxwell equations. These reduced dynamical equations describe the turbulent evolution of low-frequency electromagnetic fluctuations in nonuniform magnetized plasmas with arbitrary magnetic geometry. 
\end{abstract}

\maketitle

\section{\label{sec:intro}Introduction}

The development of gyrokinetic theory was initially motivated by the need to describe complex plasma dynamics over time scales that are long compared to the short gyration time scale of a charged particle about a nonuniform magnetic field line. Thus, gyrokinetic theory was built upon a generalization of guiding-center theory (see Refs.~\cite{Northrop} and \cite{RGL_83}). In Ref.~\cite{Taylor}, for example, Taylor showed that, while the guiding-center magnetic-moment invariant can be destroyed by low-frequency, short perpendicular-wavelength electrostatic fluctuations, a new magnetic-moment invariant can be constructed as an asymptotic expansion in powers of the amplitude of the perturbation field. This early result indicated that gyrokinetic theory was to be built upon an additional transformation beyond the guiding-center phase-space coordinates, thereby constructing new \textit{gyrocenter} phase-space coordinates, which describe gyroangle-averaged perturbed guiding-center dynamics. The purpose of the present paper is to review the foundations of gyrokinetic theory by presenting the Lagrangian and Hamiltonian methods used in the derivation of self-consistent, energy-conserving gyrokinetic Vlasov-Maxwell equations describing the nonlinear turbulent evolution of low-frequency, short perpendicular-wavelength electromagnetic fluctuations in nonuniform magnetized plasmas.

Although a comprehensive review of the applications of gyrokinetic theory merits a separate paper, we briefly point out, here, that applications are generally divided into applications based on linearized gyrokinetic equations and applications based on nonlinear gyrokinetic equations. On the one hand, linear gyrokinetic theory (see Refs.~\cite{Rutherford_Frieman}-\cite{Qin_etal}, for example) is concerned with the stability of inhomogeneous magnetized plasmas perturbed by low-frequency electromagnetic perturbations (which preserve the gyrocenter magnetic moment, or first adiabatic invariant). Nonlinear gyrokinetic theory (see Refs.~\cite{Frieman_Chen}-\cite{Hahm_96}), on the other hand, focuses its attention on the anomalous transport associated with low-frequency electromagnetic fluctuations in inhomogeneous magnetized plasmas. Gyrokinetic particle simulation techniques (see Refs.~\cite{Lee}-\cite{Parker_etal}) now play a major role in the investigation of low-frequency plasma turbulence and its associated transport in magnetized plasmas \cite{Krommes_report}.

The remaining material is organized as follows. In Sec.~\ref{sec:ext_single}, we introduce the variational principles in eight-dimensional extended phase space that are necessary to present a consistent time-dependent Hamiltonian perturbation theory for single-particle dynamics. Here, and throughout the paper, the language of differential forms is strongly emphasized within the context of Lagrangian and Hamiltonian mechanics \cite{Arnold}. In Sec.~\ref{sec:Lie}, the phase-space Lagrangian Lie-transform perturbation method, which allows a simultaneous dynamical reduction of the Hamiltonian and Poisson-bracket structure for single-particle dynamics, is briefly presented. In Sec.~\ref{sec:gyrocenter}, the phase-space Lagrangian Lie-perturbation method is used to derive gyrocenter Hamiltonian dynamics by a time-dependent phase-space transformation from guiding-center coordinates to gyrocenter coordinates. Here, we choose a Hamiltonian formulation that preserves the guiding-center Poisson-bracket structure and carries all field perturbations onto the gyrocenter Hamiltonian. 

In Sec.~\ref{sec:vp}, we show how a self-consistent set of nonlinear gyrokinetic Vlasov-Maxwell equations can be derived from a reduced variational principle. The reduced variational principle is itself shown to be derived from a variational principle for the exact Vlasov-Maxwell equations also presented in Sec.~\ref{sec:vp}. We also show how the Noether method can be used to derive exact conservation laws for the exact and gyrokinetic Vlasov-Maxwell equations. Lastly, we summarize the work presented here and discuss extensions of the standard gyrokinetic formalism in 
Sec.~\ref{sec:sum}.

\section{\label{sec:ext_single}Variational Principles for \\ Extended Single-Particle Phase-Space Dynamics}

This Section presents a brief introduction to the extended phase-space Lagrangian formulation of charged-particle dynamics in a time-dependent electromagnetic field. Here, the electromagnetic field is represented by the potentials $(\Phi,\bf{A})$, while the eight-dimensional extended phase-space noncanonical coordinates $\cal{Z} = (\bf{x},\bf{v},w,t)$ include the position $\bf{x}$ of a charged particle (mass $m$ and charge $e$), its velocity $\bf{v}$, and the canonically-conjugate time-energy $(t,w)$ coordinates. 

\subsection{Single-particle Lagrangian dynamics in extended phase space}

The phase-space Lagrangian (or Poincar\'{e}-Cartan \cite{Arnold}) one-form for a charged particle in eight-dimensional extended phase space is expressed as
\begin{equation}
\Gamma \;=\; \left( \frac{e}{c}\;\bf{A} \;+\; m\,\bf{v} \right)\bdot d\bf{x} \;-\; w\;dt \;-\; \cal{H}\;d\tau \;\equiv\; \Gamma_{a}(\cal{Z})\;
d\cal{Z}^{a} \;-\; \cal{H}(\cal{Z})\;d\tau, 
\label{eq:epsl_exact}
\end{equation}
where summation over repeated indices is implied (latin letters $a,b,c,...$ go from 1 to 8 while greek letters $\mu, \nu, ...$ go from 0 to 3), $\tau$ denotes the Hamiltonian orbit parameter, and the extended phase-space Hamiltonian is
\begin{equation}
\cal{H}(\cal{Z}) \;=\; \frac{m}{2}\;|\bf{v}|^{2} \;+\; e\;\Phi \;-\; w \;\equiv\; H(\bf{z},t) \;-\; w.
\label{eq:eHam_exact}
\end{equation}
Here, $H(\bf{z},t)$ denotes the standard time-dependent Hamiltonian, with $\bf{z} = (\bf{x},\bf{v})$, and the physical single-particle motion takes place on the subspace $\cal{H}(\cal{Z}) = H(\bf{z},t) - w = 0$ of extended phase space. 

Next, we introduce the single-particle action integral 
\[ S \;=\; \int\,\Gamma \;=\; \int \left( \Gamma_{a}\;\frac{d\cal{Z}^{a}}{d\tau} \;-\; \cal{H} \right) d\tau, \]
where $\Gamma_{a}$ are known as the \textit{symplectic} components of the phase-space Lagrangian $\Gamma$. The Principle of Least Action $\delta S = 0$ for single-particle motion in extended phase space
\begin{equation}
0 \;=\; \int\,\delta\Gamma \;=\; \int\;\delta\cal{Z}^{a} \left[\; \left( \pd{\Gamma_{b}}{\cal{Z}^{a}} \;-\; \pd{\Gamma_{a}}{\cal{Z}^{b}} \right) \; 
d\cal{Z}^{b} \;-\; \pd{\cal{H}}{\cal{Z}^{a}}\;d\tau \;\right]
\label{eq:eVP_exact}
\end{equation}
yields the extended phase-space Euler-Lagrange equations
\begin{equation}
\omega_{ab}\;\frac{d\cal{Z}^{b}}{d\tau} \;\equiv\; \left( \pd{\Gamma_{b}}{\cal{Z}^{a}} \;-\; \pd{\Gamma_{a}}{\cal{Z}^{b}} \right) 
\;\frac{d\cal{Z}^{b}}{d\tau} \;=\; \pd{\cal{H}}{\cal{Z}^{a}},
\label{eq:eEL_exact}
\end{equation}
where $\omega_{ab}$ denotes a component of the $8\times 8$ Lagrange matrix $\vb{\omega}$ \cite{Goldstein}. 

The components of the inverse of the Lagrange matrix ${\sf J} \equiv \vb{\omega}^{-1}$, known as the Poisson matrix, are the fundamental Poisson brackets 
\begin{equation}
(\vb{\omega}^{-1})^{ab} \;\equiv\; \{ \cal{Z}^{a},\; \cal{Z}^{b}\}_{\cal{Z}} \;=\; J^{ab}(\cal{Z}), 
\label{eq:epb_def}
\end{equation}
so that the Euler-Lagrange equations (\ref{eq:eEL_exact}) become the extended Hamilton's equations
\begin{equation}
\frac{d\cal{Z}^{a}}{d\tau} \;=\; J^{ab}\;\pd{\cal{H}}{\cal{Z}^{b}} \;=\; \{ \cal{Z}^{a},\; \cal{H}\}_{\cal{Z}}.
\label{eq:ehameq_exact}
\end{equation}
The fundamental Poisson brackets $J^{ab}(\cal{Z})$ satisfy the Liouville identities (which follow from the incompressibility of the extended Hamiltonian flow)
\begin{equation}
\frac{1}{\cal{J}}\;\pd{}{\cal{Z}^{a}} \left( \cal{J}\;J^{ab} \right) \;=\; 0,
\label{eq:L_id}
\end{equation}
where $\cal{J}(\cal{Z}) \equiv \sqrt{{\rm det}\,\vb{\omega}}$ defines the Jacobian associated with the choice of eight-dimensional phase-space coordinates, and the Jacobi property
\[ J^{ad}\;\pd{J^{bc}}{\cal{Z}^{d}} \;+\; J^{bd}\;\pd{J^{ca}}{\cal{Z}^{d}} \;+\; J^{cd}\;\pd{J^{ab}}{\cal{Z}^{d}} \;=\; 0, \]
obtained from the Jacobi identity 
\begin{equation}
\{f,\;\{g,\; h\}\} \;+\; \{ g,\; \{h,\; f\}\} \;+\; \{h,\; \{f,\; g\}\} \;=\; 0,  
\label{eq:J_id}
\end{equation}
which is valid for three arbitrary functions $f$, $g$, and $h$. Note that, by using the identity $\omega_{ac}\,J^{cb} = \delta_{a}^{\;b}$, the Jacobi property may be rewritten as ${\sf d}\,\omega = 0$, i.e.,
\[ \partial_{a}\omega_{bc} \;+\; \partial_{b}\omega_{ca} \;+\; \partial_{c}\omega_{ab} \;=\; 0, \]
which is always satisfied since $\omega \equiv {\sf d}\Gamma$ is an exact two-form (i.e., $\omega_{ab} = \partial_{a}\Gamma_{b} - \partial_{b}
\Gamma_{a}$). Hence, any bracket derived through the sequence $\Gamma \rightarrow \omega = {\sf d}\Gamma \rightarrow J = \omega^{-1}$ automatically satisfies the Jacobi identity (\ref{eq:J_id}).

Lastly, using the extended phase-space Lagrangian (\ref{eq:epsl_exact}), the explicit form of the extended \textit{noncanonical} Poisson bracket, defined by Eq.~(\ref{eq:epb_def}), is
\begin{eqnarray}
\{ f,\; g\}_{\cal{Z}} & = & \pd{f}{w}\,\left( \pd{g}{t} \;-\; \frac{e}{mc}\,\pd{\bf{A}}{t}\bdot\pd{g}{\bf{v}} \right)
\;-\; \left( \pd{f}{t} \;-\; \frac{e}{mc}\,\pd{\bf{A}}{t}\bdot\pd{f}{\bf{v}} \right)\,\pd{g}{w} \nonumber \\
 &  &\mbox{}+\; \frac{1}{m}\;\left( \nabla f\bdot\pd{g}{\bf{v}} - \pd{f}{\bf{v}}\bdot\nabla g \right) \;+\;
\frac{e\,\bf{B}}{m^{2}c}\bdot\pd{f}{\bf{v}}\btimes\pd{g}{\bf{v}}.
\label{eq:ePB_exact}
\end{eqnarray}
Here, the noncanonical aspect of the Poisson bracket (\ref{eq:ePB_exact}) is exhibited by the appearance of space-time derivatives of the vector potential $\bf{A}(\bf{x},t)$. The Hamiltonian dynamics in extended phase space is expressed in terms of Eqs.~(\ref{eq:ehameq_exact})-(\ref{eq:ePB_exact}) as
\begin{eqnarray*}
\frac{d\bf{x}}{d\tau} & = & \frac{1}{m}\;\pd{\cal{H}}{\bf{v}} \;=\; \bf{v}, \\
\frac{d\bf{v}}{d\tau} & = & -\,\frac{1}{m}\;\nabla\cal{H} \;+\; \frac{e}{mc}\, \left( \pd{\bf{A}}{t}\;\pd{\cal{H}}{w} \;+\; 
\pd{\cal{H}}{\bf{v}}\btimes\frac{\bf{B}}{m} \right) \;=\; \frac{e}{m} \left( \bf{E} \;+\; \frac{\bf{v}}{c}\btimes\bf{B} \right), \\
\frac{dw}{d\tau} & = & \pd{\cal{H}}{t} \;-\; \frac{e}{mc}\,\pd{\bf{A}}{t}\bdot\pd{\cal{H}}{\bf{v}} \;=\; e \left(
\pd{\Phi}{t} \;-\; \frac{\bf{v}}{c}\bdot\pd{\bf{A}}{t} \right), \\
\frac{dt}{d\tau} & = & -\;\pd{\cal{H}}{w} \;=\; 1.
\end{eqnarray*}
Note that the Hamiltonian orbit parameter $\tau$ can be identified with the time coordinate $t$ and, as expected, the energy coordinate is an invariant when the electromagnetic fields are time-independent.

\subsection{Single-particle perturbation theory in extended phase space}

It turns out that the eight-dimensional extended phase space provides a natural setting for time-dependent Hamiltonian perturbation theory.
A variational formulation of single-particle perturbation theory can be introduced through the new phase-space Lagrangian one-form \cite{Brizard_01}
\begin{equation}
\Gamma^{\prime} \;\equiv\; \Gamma_{a}\;d\cal{Z}^{a} \;-\; \cal{H}\;d\tau \;-\; \cal{S}\;d\epsilon,
\label{eq:epsl_pert}
\end{equation}
where the symplectic components $\Gamma_{a}$ and the Hamiltonian $\cal{H}$ now depend on the perturbation parameter $\epsilon$ and the scalar field 
$\cal{S}$ is the generating function for an infinitesimal canonical transformation that smoothly deforms a particle's extended phase-space orbit from a reference orbit (at $\epsilon = 0$) to a perturbed orbit (for $\epsilon \neq 0$). From the phase-space Lagrangian (\ref{eq:epsl_pert}), we construct the action \textit{path}-integral $S_{C}^{\prime} = \int_{C}\,\Gamma^{\prime}$ evaluated along a fixed path $C$ in the $(\tau,\epsilon)$-parameter space.

The modified Principle of Least Action for perturbed single-particle motion in extended phase space
\begin{equation}
0 \;=\; \int\,\delta\Gamma^{\prime} \;=\; \int\;\delta\cal{Z}^{a} \left[\; \omega_{ab}\; d\cal{Z}^{b} \;-\; \pd{\cal{H}}{\cal{Z}^{a}}\;d\tau \;-\; 
\left( \pd{\cal{S}}{\cal{Z}^{a}} \;+\; \pd{\Gamma_{a}}{\epsilon} \right) d\epsilon \;\right]
\label{eq:eVP_pert}
\end{equation}
now yields the extended perturbed Hamilton's equations
\begin{eqnarray}
\frac{d\cal{Z}^{a}}{d\tau} & = & \left\{ \cal{Z}^{a},\; \cal{H}\right\}_{\cal{Z}}, \label{eq:eHam_H} \\
\frac{d\cal{Z}^{a}}{d\epsilon} & = & \left\{ \cal{Z}^{a},\; \cal{S}\right\}_{\cal{Z}} \;-\;
\pd{\Gamma_{b}}{\epsilon}\;\left\{\cal{Z}^{b},\; \cal{Z}^{a} \right\}_{\cal{Z}}, \label{eq:eHam_S}
\end{eqnarray}
where Eq.~(\ref{eq:eHam_H}) is identical to Eq.~(\ref{eq:ehameq_exact}) except that the extended Hamiltonian and Poisson bracket now depend on the perturbation parameter $\epsilon$, while Eq.~(\ref{eq:eHam_S}) determines how particle orbits evolve under the perturbation $\epsilon$-flow. 

We note that the order of time evolution ($\tau$-flow) and perturbation evolution ($\epsilon$-flow) should be immaterial (i.e., we may evolve the system along its reference orbit and then perturb it its final state, or perturb the initial state and evolve the perturbed orbit to the same final state). The commutativity of the two Hamiltonian $(\tau,\,\epsilon)$ flows, therefore, leads to the path independence of the action integral $\int\Gamma^{\prime}$ in the two-dimensional $(\tau,\epsilon)$ orbit-parameter space. Thus, considering two arbitrary paths $C$ and $\ov{C}$ with identical end points on the $(\tau,\,\epsilon)$-parameter space and calculating the action path-integrals $S_{C}^{\prime} = \int_{C}\,\Gamma^{\prime}$ and $S_{\ov{C}}^{\prime} = \int_{\ov{C}}\,\Gamma^{\prime}$, the path-independence condition $(S_{\ov{C}}^{\prime} = S_{C}^{\prime})$ leads (by Stokes' Theorem for differential one-forms \cite{Spivak}) to the condition 
\[ 0 \;=\; \int_{C}\,\Gamma^{\prime} \;-\; \int_{\ov{C}}\,\Gamma^{\prime} \;=\; \oint_{\partial D}\;\Gamma^{\prime} \;=\; \int_{D}\,d\Gamma^{\prime}, \] where $D$ is the area enclosed by the closed path $\partial D \equiv C - \ov{C}$, and the two-form $d\Gamma^{\prime}$ on the $(\tau,\,\epsilon)$-parameter space is
\begin{eqnarray*} 
d\Gamma^{\prime} & = & d\epsilon\wedge d\tau \left[ \frac{d\cal{Z}^{a}}{d\epsilon}\,\omega_{ab}\,\frac{d\cal{Z}^{b}}{d\tau} \;-\;
\left( \pd{\cal{H}}{\epsilon} + \pd{\cal{H}}{\cal{Z}^{a}}\,\frac{d\cal{Z}^{a}}{d\epsilon} \right) \;+\; \left( \pd{\cal{S}}{\cal{Z}^{a}} + 
\pd{\Gamma_{a}}{\epsilon}\right) \frac{d\cal{Z}^{a}}{d\tau} \right] \\
 & \equiv & d\epsilon\wedge d\tau \left( \{ \cal{S},\;\cal{H}\}_{\cal{Z}} \;-\; \pd{\cal{H}}{\epsilon} \;+\; \pd{\Gamma_{a}}{\epsilon}\;\left\{ 
\cal{Z}^{a},\;\cal{H} \right\}_{\cal{Z}} \right),
\end{eqnarray*}
where Eqs.~(\ref{eq:eHam_H})-(\ref{eq:eHam_S}) were used. Hence, the condition of path independence requires that $d\Gamma^{\prime} = 0$, which yields the Hamiltonian perturbation equation
\begin{equation}
\{\cal{S},\; \cal{H} \}_{\cal{Z}} \;\equiv\; \pd{\cal{H}}{\epsilon} \;-\; \pd{\Gamma_{a}}{\epsilon}\;
\left\{ \cal{Z}^{a},\;\cal{H} \right\}_{\cal{Z}},
\label{eq:eHam_perteq}
\end{equation}
relating the generating scalar field $\cal{S}$ to the perturbation-parameter dependence of the extended Hamiltonian $(\partial_{\epsilon}\cal{H})$ and Poisson bracket $(\partial_{\epsilon}\Gamma_{a})$. Using the extended single-particle Hamiltonian (\ref{eq:eHam_exact}), the Hamiltonian perturbation equation (\ref{eq:eHam_perteq}) becomes
\begin{equation}
\{\cal{S},\; \cal{H}\}_{\cal{Z}} \;=\; e\;\pd{\Phi}{\epsilon} \;-\; \frac{e}{c}\;\pd{\bf{A}}{\epsilon}\bdot
\left\{ \bf{x},\;\cal{H} \right\}_{\cal{Z}},
\label{eq:electro_perteq}
\end{equation}
whose formal solution is
\[ \cal{S} \;\equiv\; \left( \frac{d}{d\tau}\right)^{-1} \left[\; e\;\pd{\Phi}{\epsilon} \;-\; \frac{e}{c}\;\pd{\bf{A}}{\epsilon}\bdot\left\{ \bf{x},\;
\cal{H} \right\}_{\cal{Z}} \;\right], \]
where $(d/d\tau)^{-1}$ denotes an integration along a perturbed Hamiltonian orbit. Here, the perturbed evolution operator $d/d\tau$ is expanded in powers of $\epsilon$, with the lower-order operator $d_{0}/d\tau$ considered to be explicitly integrable. In practice, the generating function $\cal{S}$ is also expanded in powers of $\epsilon$: $\cal{S} = \cal{S}_{1} + \epsilon\,\cal{S}_{2} + \cdots$, so that the first-order term is expressed as
\begin{equation}
\cal{S}_{1} \;\equiv\; \left( \frac{d_{0}}{d\tau}\right)^{-1} \left[\; e\;\Phi_{1} \;-\; e\;\bf{A}_{1}\bdot\frac{\bf{v}_{0}}{c} \;\right],
\label{eq:S_formal}
\end{equation}
where $\bf{v}_{0} \equiv d_{0}\bf{x}/d\tau = \{ \bf{x},\;\cal{H}_{0}\}_{0}$ denotes the particle's unperturbed velocity. In order to determine the higher-order terms $\cal{S}_{n}$ (for $n \geq 2$), a more systematic approach, based on applications of the Lie-transform perturbation method, is required.

\section{\label{sec:Lie}Lie-transform Perturbation Theory}

The Hamiltonian perturbation equation (\ref{eq:eHam_perteq}) arises naturally within the context of the dynamical reduction of single-particle Hamilton's equations (\ref{eq:eHam_H}) through the elimination of fast \textit{orbital} time scales. The most efficient method for deriving reduced Hamilton's equations is based on Hamiltonian \cite{osc} and phase-space Lagrangian \cite{Cary_RGL} Lie-transform perturbation methods. 

\subsection{Near-identity phase-space transformations}

The process by which a fast time scale is removed from Hamilton's equations $\{\cal{Z}^{a},\; \cal{H}\}_{\cal{Z}}$ involves a near-identity transformation on extended particle phase space \cite{RGL_82}: 
\begin{equation}
\cal{T}_{\epsilon}:\; \cal{Z} \;\rightarrow\; \ov{\cal{Z}}(\cal{Z};\epsilon) \;\equiv\; \cal{T}_{\epsilon}\cal{Z},\;\; 
{\rm with}\;\; \ov{\cal{Z}}(\cal{Z};0) \;=\; \cal{Z},
\label{eq:near_trans}
\end{equation}
where $\epsilon \ll 1$ denotes a dimensionless ordering parameter. Here, the near-identity transformation is explicitly expressed in terms of generating vector fields $(\cal{G}_{1},\cal{G}_{2},...)$:
\begin{equation}
\ov{\cal{Z}}^{a}(\cal{Z},\epsilon) \;=\; \cal{Z}^{a} \;+\; \epsilon\,\cal{G}_{1}^{a} \;+\; \epsilon^{2} \left( 
\cal{G}_{2}^{a} \;+\; \frac{1}{2}\, \cal{G}_{1}^{b}\;\pd{\cal{G}_{1}^{a}}{\cal{Z}^{b}} \right) \;+\; \cdots,
\label{eq:ovZ_Z}
\end{equation}
where the $n$th-order generating vector field $\cal{G}_{n}$ is chosen to remove the fast time scale at order $\epsilon^{n}$ from the perturbed Hamiltonian dynamics. The new extended phase-space coordinates include the pair of fast action-angle coordinates $(\ov{J},\ov{\theta})$ and the reduced phase-space coordinates $\ov{\cal{Z}}_{R}$ such that the fast action $\ov{J} = \ov{J}_{0} + \epsilon\,\ov{J}_{1} + \cdots$ is an exact invariant of the reduced Hamiltonian dynamics and the Hamiltonian dynamics of the reduced coordinates $\ov{\cal{Z}}_{R}$ is independent of the fast angle $\ov{\theta}$. The small dimensionless parameter can, therefore, be defined as $\epsilon \equiv (\tau_{R}\,\Omega)^{-1} \ll 1$, where $\tau_{R}$ is the characteristic time scale of the reduced dynamics and $(2\pi/\Omega)$ denotes the fast orbital time scale associated with the fast angle $\ov{\theta}$.

Next, using the transformation (\ref{eq:near_trans}), we define the \textit{push-forward} operator on scalar fields
\cite{RGL_82} \textit{induced} by the near-identity transformation (\ref{eq:near_trans}):
\begin{equation}
{\sf T}_{\epsilon}^{-1}:\; \cal{F} \;\rightarrow\; \ov{\cal{F}} \;\equiv\; {\sf T}_{\epsilon}^{-1}\cal{F},
\label{eq:pushforward}
\end{equation}
i.e., ${\sf T}_{\epsilon}^{-1}$ transforms a scalar field $\cal{F}$ on the phase space with coordinates $\cal{Z}$ into a scalar field $\ov{\cal{F}}$ on the phase space with coordinates $\ov{\cal{Z}}$: 
\[ \ov{\cal{F}}(\ov{\cal{Z}}) \;=\; {\sf T}_{\epsilon}^{-1}\cal{F}(\ov{\cal{Z}}) \;=\; \cal{F}(\cal{T}_{\epsilon}^{-1}
\ov{\cal{Z}}) \;=\; \cal{F}(\cal{Z}). \]
Since the transformation (\ref{eq:near_trans}) is invertible, i.e., there exists an inverse near-identity transformation 
\begin{equation}
\cal{T}_{\epsilon}^{-1}:\; \ov{\cal{Z}} \;\rightarrow\; \cal{Z}(\ov{\cal{Z}};\epsilon) \;\equiv\; \cal{T}_{\epsilon}^{-1}
\ov{\cal{Z}},\;\; {\rm with}\;\; \cal{Z}(\ov{\cal{Z}};0) \;=\; \ov{\cal{Z}},
\label{eq:near_inv}
\end{equation}
we also define the {\it pull-back} operator \cite{RGL_82}: 
\begin{equation}
{\sf T}_{\epsilon}:\; \ov{\cal{F}} \;\rightarrow\; \cal{F} \;\equiv\; {\sf T}_{\epsilon}\ov{\cal{F}},
\label{eq:pullback}
\end{equation}
i.e., ${\sf T}_{\epsilon}$ transforms a scalar field $\ov{\cal{F}}$ on the phase space with coordinates $\ov{\cal{Z}}$ into a scalar field $\cal{F}$ on the phase space with coordinates $\cal{Z}$: 
\[ \cal{F}(\cal{Z}) \;=\; {\sf T}_{\epsilon}\ov{\cal{F}}(\cal{Z}) \;=\; \ov{\cal{F}}({\cal T}_{\epsilon}\cal{Z}) \;=\; 
\ov{\cal{F}}(\ov{\cal{Z}}). \]
Using the fact that the total $\tau$-derivative of a scalar field $\cal{F}$ is itself a scalar field, we obtain an expression for the transformed
operator $d_{\epsilon}/d\tau$ defined as
\begin{equation}
\frac{d_{\epsilon}\ov{\cal{F}}}{d\tau} \;\equiv\; {\sf T}_{\epsilon}^{-1}\left( \frac{d}{d\tau}\;{\sf T}_{\epsilon}\ov{\cal{F}} \right) \;=\; 
\{ \ov{\cal{F}},\; \ov{\cal{H}} \}_{\ov{\cal{Z}}},
\label{eq:PB_trans}
\end{equation} 
where $\{\;,\;\}_{\ov{\cal{Z}}}$ denotes the new transformed Poisson bracket and
\begin{equation}
\ov{\cal{H}} \;\equiv\; {\sf T}_{\epsilon}^{-1}\cal{H}
\label{eq:eham_lt}
\end{equation}
denotes the transformed Hamiltonian. Once again, the new extended phase-space coordinates are chosen so that $d_{\epsilon}\ov{J}/d\tau = \{\ov{J},\; 
\ov{\cal{H}}\}_{\ov{\cal{Z}}} \equiv 0$ and $d_{\epsilon}\ov{Z}_{R}/d\tau = \{\ov{Z}_{R},\; \ov{\cal{H}}\}_{\ov{\cal{Z}}}$ be independent of the fast angle $\ov{\theta}$. The dynamical reduction of single-particle Hamiltonian dynamics consists in the construction of a fast invariant $\ov{J}$ with its canonically-conjugate fast-angle $\ov{\theta}$ becoming an ignorable coordinate.

\subsection{Lie derivatives and Lie transforms}

In Lie-transform perturbation theory \cite{RGL_82}, the push-forward operator (\ref{eq:pushforward}) is defined as
\begin{equation} 
{\sf T}_{\epsilon}^{-1} \;\equiv\; \cdots\;\;\exp\left(-\;\epsilon^{2}\,L_{2}\right)\;\exp\left(-\;\epsilon\,L_{1}\right)
\label{eq:lt_def}
\end{equation}
in terms of the $n$th-order Lie derivative $L_{n}$ generated by the $n$th-order vector field $\cal{G}_{n}$ \cite{Abraham_Marsden}. In 
Eq.~(\ref{eq:eham_lt}), the Lie derivative $L_{n}\cal{H}$ is defined as the scalar field 
\begin{equation}
L_{n}\cal{H} \;\equiv\; \cal{G}_{n}^{a}\,\partial_{a}\cal{H}. 
\label{eq:Lie_zero}
\end{equation}
The transformation of the Poisson bracket by Lie-transform methods, on the other hand, is performed through the transformation of the extended phase-space Lagrangian, expressed as 
\begin{equation}
\ov{\Gamma} \;=\; {\sf T}_{\epsilon}^{-1}\Gamma \;+\; {\sf d}\cal{S}, 
\label{eq:epsl_lt}
\end{equation}
where $\cal{S}$ denotes a (gauge) scalar field used to simplify the transformed Hamiltonian (\ref{eq:eham_lt}), i.e., it has no impact on the new
Poisson-bracket structure $\ov{\omega} = {\sf d}\ov{\Gamma} = {\sf d}({\sf T}_{\epsilon}^{-1}\Gamma)$ since ${\sf d}^{2}\cal{S} = 0$ (i.e., 
$\partial^{2}_{ab}\cal{S} - \partial^{2}_{ba}\cal{S} = 0$). In Eq.~(\ref{eq:epsl_lt}), the $n$th-order Lie derivative $L_{n}\Gamma$ of a one-form $\Gamma \equiv \Gamma_{a}\,{\sf d}\cal{Z}^{a}$ is defined as a one-form \cite{Abraham_Marsden}
\begin{equation} 
L_{n}\Gamma \;\equiv\; \cal{G}_{n}^{a}\;\omega_{ab}\;{\sf d}\cal{Z}^{b} \;+\; {\sf d}\left( \cal{G}_{n}^{a}\;\Gamma_{a}\right),
\label{eq:Lie_one}
\end{equation}
where, at each order, the terms ${\sf d}(\cdots)$ can be absorbed in the gauge term ${\sf d}\cal{S}_{n}$.

\subsubsection{Transformed extended Poisson-bracket structure}

We now write the extended phase-space Lagrangian $\Gamma \equiv \Gamma_{0} + \epsilon\,\Gamma_{1}$ and the extended Hamiltonian $\cal{H} \equiv 
\cal{H}_{0} + \epsilon\,\cal{H}_{1}$ in terms of an unperturbed (zeroth-order) part and a perturbation (first-order) part. The Lie-transform relations associated with Eq.~(\ref{eq:epsl_lt}) are expressed (up to second order in $\epsilon$) as
\begin{eqnarray}
\ov{\Gamma}_{0a} & = & \Gamma_{0a}, \label{eq:epsl_0} \\
\ov{\Gamma}_{1a} & = & \Gamma_{1a} \;-\; \cal{G}_{1}^{b}\;\omega_{0ba} \;+\; \partial_{a}\cal{S}_{1}, 
\label{eq:epsl_1} \\
\ov{\Gamma}_{2a} & = & -\; \cal{G}_{2}^{b}\;\omega_{0ba} \;-\; \frac{1}{2}\;\cal{G}_{1}^{b} \left(\omega_{1ba} \;+\;
\ov{\omega}_{1ba} \right) \;+\; \partial_{a}\cal{S}_{2}, \label{eq:epsl_2}
\end{eqnarray}
where each scalar field $\cal{S}_{n}$ $(n \geq 1)$ is determined by requiring that the $n$th-order Hamiltonian $\ov{\cal{H}}_{n}$ be independent of the fast orbital time scale. The simplest form for the new Poisson bracket $\{\;,\;\}_{\ov{\cal{Z}}}$ is obtained by choosing $\ov{\Gamma} \equiv 
\ov{\Gamma}_{0} = \Gamma_{0}$, so that the condition $\ov{\Gamma}_{n} \equiv 0$ $(n \geq 1)$ yield a solution for the generating vector field 
$\cal{G}_{n}$ expressed in terms of the scalar fields $(\cal{S}_{1},\cdots,\cal{S}_{n})$. 

For the first-order generating vector field $\cal{G}_{1}$, the condition $\ov{\Gamma}_{1} \equiv 0$ yields the following expression in terms of the scalar field $\cal{S}_{1}$:
\begin{equation}
\cal{G}_{1}^{a} \;=\; \left\{ \cal{S}_{1},\; \cal{Z}^{a}\right\}_{0} \;+\; \Gamma_{1b}\;J_{0}^{ba},
\label{eq:G1_def}
\end{equation}
where $J_{0}^{ab}$ denotes a component of the zeroth-order Poisson matrix. Next, for the second-order generating vector field $\cal{G}_{2}$, the condition $\ov{\Gamma}_{2} \equiv 0$ yields the following expression in terms of the scalar field $\cal{S}_{2}$:
\begin{equation}
\cal{G}_{2}^{a} \;=\; \left\{ \cal{S}_{2},\; \cal{Z}^{a}\right\}_{0} \;-\; \frac{1}{2}\;\cal{G}_{1}^{b}\;\omega_{1\,bc}\;J_{0}^{ca},
\label{eq:G2_def}
\end{equation}
where $\omega_{1\,bc}$ denotes the component of the first-order perturbed Lagrange matrix. The near-identity extended phase-space transformation 
(\ref{eq:ovZ_Z}) is, thus, expressed (up to first order in $\epsilon$) as
\begin{equation}
\ov{\cal{Z}}^{a} \;=\; \cal{Z}^{a} \;+\; \epsilon \left( \left\{ \cal{S}_{1},\; \cal{Z}^{a} \right\}_{0} \;+\; \Gamma_{1b}\;J_{0}^{ba} \right) \;+\;
\cdots,
\label{eq:Lie_near}
\end{equation}
and its explicit expression requires a solution of the scalar fields $(\cal{S}_{1},\cdots)$.

\subsubsection{Transformed extended Hamiltonian}

By substituting the generating vector fields (\ref{eq:G1_def}) and (\ref{eq:G2_def}) into the Lie-transform relations associated with 
Eq.~(\ref{eq:eham_lt}):
\begin{eqnarray*}
\ov{\cal{H}}_{0} & = & \cal{H}_{0}, \\
\ov{\cal{H}}_{1} & = & \cal{H}_{1} \;-\; \cal{G}_{1}^{a}\,\partial_{a}\cal{H}_{0}, \\
\ov{\cal{H}}_{2} & = & -\; \cal{G}_{2}^{a}\,\partial_{a}\cal{H}_{0} \;-\; \frac{1}{2}\;\cal{G}_{1}^{a}\;\partial_{a}\left( \cal{H}_{1} 
\;+\; \ov{\cal{H}}_{1} \right),
\end{eqnarray*}
we obtain the first-order and second-order terms in the transformed extended Hamiltonian:
\begin{equation}
\ov{\cal{H}}_{1} \;=\; \cal{H}_{1} \;-\; \Gamma_{1\,a}\;\left\{ \cal{Z}^{a},\; \cal{H}_{0}\right\}_{0} \;-\; \left\{ \cal{S}_{1},\; \cal{H}_{0} 
\right\}_{0} \;\equiv\; \cal{K}_{1} \;-\; \left\{ \cal{S}_{1},\; \cal{H}_{0} \right\}_{0}, 
\label{eq:eham_1}
\end{equation}
where $\cal{K}_{1} \equiv \cal{H}_{1} - \Gamma_{1\,a}\,\{ \cal{Z}^{a},\; \cal{H}_{0}\}_{0}$ denotes the \textit{effective} first-order Hamiltonian, and
\begin{eqnarray}
\ov{\cal{H}}_{2} & = & -\; \left\{ \cal{S}_{2},\; \cal{H}_{0} \right\}_{0} \;-\; \left. \left. \frac{1}{2} \right( \left\{ \cal{S}_{1},\; \cal{H}_{1} + 
\ov{\cal{H}}_{1} \right\}_{0} \;+\; \Gamma_{1a}\; \left\{ \cal{Z}^{a},\; \cal{H}_{1} + \ov{\cal{H}}_{1}\right\}_{0} \right) \nonumber \\
 &  &\mbox{}+\; \frac{1}{2}\;\cal{G}_{1}^{a}\;\omega_{1ab}\;\left\{ \cal{Z}^{b},\; \cal{H}_{0}\right\}_{0}. \label{eq:eham_2}
\end{eqnarray}
The solution for the new first-order Hamiltonian (\ref{eq:eham_1}) is expressed in terms of the fast-angle averaging operation $\langle\cdots\rangle$ as
\begin{equation}
\ov{\cal{H}}_{1} \;\equiv\; \langle\cal{K}_{1}\rangle \;=\; \langle\cal{H}_{1}\rangle \;-\; \left\langle \Gamma_{1\,a}\;\left\{ \cal{Z}^{a},\; 
\cal{H}_{0}\right\} \right\rangle,
\label{eq:H1_average}
\end{equation}
where the Poisson bracket $\{\;,\;\}$ is the zeroth-order Poisson bracket $\{\;,\;\}_{0}$ (unless otherwise noted) and $\cal{S}_{1}$ can be chosen such that $\langle\cal{S}_{1}\rangle \equiv 0$. The first-order scalar field $\cal{S}_{1}$ is, thus, the solution to the perturbation equation
\begin{equation}
\{ \cal{S}_{1},\; \cal{H}_{0}\} \;=\; \wt{\cal{K}}_{1} \;\equiv\; \cal{K}_{1} \;-\; \langle\cal{K}_{1}\rangle \;\;\;\rightarrow\;\;\;
\cal{S}_{1} \equiv (d_{0}/d\tau)^{-1}\wt{\cal{K}}_{1},
\label{eq:S1_tilde}
\end{equation}
where $(d_{0}/d\tau)^{-1}$ denotes an integration along an unperturbed extended Hamiltonian orbit; note the similarity of Eq.~(\ref{eq:S1_tilde}) with 
Eq.~(\ref{eq:S_formal}). To lowest order in the fast orbital time scale, the unperturbed integration $(d_{0}/d\tau)^{-1}\wt{\cal{K}}_{1} \equiv 
\Omega^{-1}\;\int \wt{\cal{K}}_{1} d\ov{\theta}$ involves an indefinite fast-angle integration.

The solution for the new second-order Hamiltonian (\ref{eq:eham_2}) can be simplified as follows. First, we introduce the Poisson-bracket identity (valid for four arbitrary functions $f$, $g$, $h$, and $k$)
\[ \{f,\;g\}\;\{h,\;k\} - \{f,\; h\}\;\{g,\; k\} = \{f,\; (g\,\{h,\; k\})\} \;+\; \{(g\,\{h,\;f\}),\; k\} \;-\; g\,\{h,\; \{f,\; k\}\}, \]
which follows from the Jacobi identity (\ref{eq:J_id}), so that the last term in Eq.~(\ref{eq:eham_2})
\begin{eqnarray*} 
\cal{G}_{1}^{a}\;\omega_{1ab}\;\left\{ \cal{Z}^{b},\; \cal{H}_{0}\right\} & = & \left( \left\{ \cal{S}_{1},\; \Gamma_{1a}\right\}\;\left\{ 
\cal{Z}^{a},\; \cal{H}_{0}\right\} \;-\; \left\{ \cal{S}_{1},\; \cal{Z}^{a} \right\}\; \left\{ \Gamma_{1a},\; \cal{H}_{0}\right\} \right) \\
 & + & \Gamma_{1a} \left( \left\{ \cal{Z}^{a},\; \Gamma_{1b}\right\}\;\left\{ \cal{Z}^{b},\; \cal{H}
\right\} \;-\; \left\{ \cal{Z}^{a},\; \cal{Z}^{b} \right\}\; \left\{ \Gamma_{1b},\; \cal{H}_{0}\right\} \right)
\end{eqnarray*}
can be rearranged and inserted back into Eq.~(\ref{eq:eham_2}) to yield
\begin{eqnarray}
\ov{\cal{H}}_{2} & = & -\; \left\{ \left( \cal{S}_{2} \;-\; \frac{1}{2}\;\Gamma_{1a}\,\{\cal{Z}^{a},\; \cal{S}_{1}\} \right),\; \cal{H}_{0}
\right\} \;-\; \left\{ \cal{S}_{1},\; \left( \langle \cal{K}_{1}\rangle \;+\; \frac{1}{2}\; \wt{\cal{K}}_{1} \right)\right\} \nonumber \\
 &  &\mbox{}-\; \Gamma_{1a} \left( \{\cal{Z}^{a},\; \cal{K}_{1}\} \;+\; \left. \left. \frac{1}{2}\;\Gamma_{1b}\,\right\{ \cal{Z}^{b},\; \{\cal{Z}^{a},\; \cal{H}_{0}\} \right\} \right).
\label{eq:eham2_new}
\end{eqnarray}
Since $\cal{S}_{2}$ can also be chosen such that $\langle\cal{S}_{2}\rangle \equiv 0$, the fast-angle average of the right side of Eq.~(\ref{eq:eham2_new})
yields the new second-order Hamiltonian
\begin{eqnarray}
\ov{\cal{H}}_{2} & = & \left. \left. \frac{1}{2}\;\right\langle \Gamma_{1a}\;\left\{\cal{Z}^{a},\; \{\cal{Z}^{b},\; \cal{H}_{0}\} \right\}
\;\Gamma_{1b} \right\rangle \;-\; \left. \left. \frac{1}{2}\; \right\langle \left\{ \cal{S}_{1},\; \wt{\cal{K}}_{1}\right\}\right\rangle \nonumber \\
 &  &\mbox{}-\; \left\langle \Gamma_{1a} \left( \{ \cal{Z}^{a},\; \cal{H}_{1} \} \;-\; \{\cal{Z}^{a},\; \Gamma_{1b}\}\;\{\cal{Z}^{b},\;
\cal{H}_{0} \} \right) \right\rangle. 
\label{eq:H2_average}
\end{eqnarray}
In the next Section, we will show that the first term in Eq.~(\ref{eq:H2_average}) corresponds to the expected quadratic nonlinearity associated with a perturbed canonical representation in which only the perturbation vector potential 
$\bf{A}_{1}$ appears in the expression for the reduced momentum coordinates. The second term, on the other hand, corresponds to the low-frequency limit of the standard quadratic ponderomotive Hamiltonian \cite{osc}. Lastly, the third set of terms will be shown to vanish because the Poisson brackets $\{\bf{x},\,\Phi_{1}(\bf{x})\}$ and $\{\bf{x},\,
\bf{A}_{1}(\bf{x})\}$ both vanish.

\section{\label{sec:gyrocenter}Nonlinear Low-frequency Gyrocenter Hamiltonian Dynamics}

In this Section, we apply the Lie-transform perturbation methods to the dynamical reduction of the perturbed dynamics of charged particles (mass $m$ and charge $e$) moving in a background time-independent magnetic field $\bf{B}_{0} = \nabla\btimes\bf{A}_{0}$ in the presence of low-frequency electromagnetic fluctuations represented by the perturbation four-potential $A_{1}^{\mu} = (\Phi_{1},\bf{A}_{1})$, whose amplitude is ordered with a dimensionless small parameter $\epsilon_{\delta} \ll 1$. 

The eight-dimensional extended phase-space dynamics is expressed in terms of the extended phase-space Lagrangian $\Gamma = \Gamma_{0} + \epsilon_{\delta}\,\Gamma_{1}$, where $\Gamma_{0} \equiv [(e/c)\,\bf{A}_{0} + m\bf{v}]\cdot d\bf{x} - w\,dt$ and $\Gamma_{1} \equiv (e/c)\, \bf{A}_{1}\bdot d\bf{x}$, and the extended phase-space Hamiltonian $\cal{H} = \cal{H}_{0} + \epsilon_{\delta}\,\cal{H}_{1}$, where $\cal{H}_{0} \equiv (m/2)\,|\bf{v}|^{2} - w$ and 
$\cal{H}_{1} \equiv e\,\Phi_{1}$. The extended Poisson bracket $\{\;,\;\}_{\cal{Z}}$ is obtained from the extended phase-space Lagrangian $\Gamma$ by standard means \cite{RGL_82}, as described earlier. Note that, while electrostatic fluctuations perturb the Hamiltonian alone, full electromagnetic fluctuations perturb both the Hamiltonian and the Poisson bracket.

The standard gyrokinetic analysis for magnetized plasmas perturbed by low-frequency electromagnetic fluctuations \cite{Brizard_89} proceeds by a 
sequence of two near-identity phase-space transformations: a time-independent {\it guiding-center} phase-space transformation and a time-dependent 
{\it gyrocenter} phase-space transformation. 

\subsection{Unperturbed guiding-center Hamiltonian dynamics}

The {\it guiding-center} phase-space transformation involves an asymptotic expansion, with a small dimensionless parameter $\epsilon_{B} \equiv\rho/L_{B} 
\ll 1$ defined as the ratio of the characteristic gyroradius $\rho$ and the background magnetic-field length scale $L_{B}$. This transformation is designed to remove the fast gyromotion time scale associated with the time-independent background magnetic field $\bf{B}_{0}$ associated with an unperturbed magnetized plasma \cite{RGL_83}. In previous work \cite{Brizard_95}, this transformation was carried out to second order in $\epsilon_{B}$ with the scalar potential $\Phi_{0}$ ordered at zeroth order in $\epsilon_{B}$. 

The results of the guiding-center analysis presented in Ref.~\cite{RGL_83} are summarized as follows. First, the guiding-center transformation yields the following guiding-center coordinates $(\bf{R},p_{\|},\mu,\theta,w,t) \equiv \cal{Z}_{gc}$, where $\bf{R}$ is the guiding-center position, $p_{\|}$ is the guiding-center kinetic momentum parallel to the unperturbed magnetic field, $\mu$ is the guiding-center magnetic moment, $\theta$ is the gyroangle, and $(w,t)$ are the canonically conjugate guiding-center energy-time coordinates (here, time is unaffected by the transformation while the guiding-center energy is chosen to be equal to the particle energy). Next, the unperturbed guiding-center extended phase-space Lagrangian is 
\begin{equation}
\Gamma_{gc} \;\equiv\; \frac{e}{c}\,\bf{A}_{0}^{*} \bdot d\bf{R} \;+\; \mu\;(mc/e)\,d\theta \;-\; w\; dt, 
\label{eq:gc_epsl0}
\end{equation}
where $\bf{A}_{0}^{*} \equiv \bf{A}_{0} + (c/e)\,p_{\|}\,\wh{\bf{b}}_{0}$ is the effective unperturbed vector potential, with $\wh{\bf{b}}_{0} \equiv 
\bf{B}_{0}/B_{0}$; we, henceforth, omit displaying the dimensionless guiding-center parameter $\epsilon_{B}$ for simplicity. The unperturbed extended phase-space guiding-center Hamiltonian is
\begin{equation}
\cal{H}_{gc} \;=\; \frac{p_{\|}^{2}}{2m} \;+\; \mu\,B_{0} \;-\; w \;\equiv\; H_{gc} - w.
\label{eq:gc_eH0}
\end{equation}
Lastly, from the unperturbed guiding-center phase-space Lagrangian (\ref{eq:gc_epsl0}), we obtain the unperturbed guiding-center Poisson bracket 
$\{\;,\;\}_{gc}$, given here in terms of two arbitrary functions $\cal{F}$ and $\cal{G}$ on extended guiding-center phase space as \cite{RGL_83}
\begin{eqnarray}
\left\{ \cal{F},\, \cal{G} \right\}_{\cal{Z}} & \equiv & \frac{e}{mc}\, \left( \pd{\cal{F}}{\theta}\, \pd{\cal{G}}{\mu} - 
\pd{\cal{F}}{\mu}\, \pd{\cal{G}}{\theta} \right) \;+\; \frac{\bf{B}_{0}^{*}}{B_{0 \|}^{*}}\bdot \left( \nabla \cal{F}\,
\pd{\cal{G}}{p_{\|}} - \pd{\cal{F}}{p_{\|}}\, \nabla\cal{G} \right) \nonumber \\
 &   &\mbox{}-\; \frac{c\wh{\bf{b}}_{0}}{eB_{0 \|}^{*}}\bdot \nabla \cal{F}\btimes\nabla \cal{G} \;+\; \left( \pd{\cal{F}}{w}\, 
\pd{\cal{G}}{t} - \pd{\cal{F}}{t}\, \pd{\cal{G}}{w} \right),
\label{eq:gc_ePB}
\end{eqnarray}
where $\bf{B}_{0}^{*} \equiv \nabla\btimes\bf{A}_{0}^{*}$ and $B_{0 \|}^{*} \equiv \wh{\bf{b}}_{0}\bdot\bf{B}_{0}^{*} = B_{)} + (c/e)p_{\|} \bhat_{0}
\bdot\nabla\btimes\bhat_{0}$; note that the Jacobian of the guiding-center transformation is $\cal{J}_{gc} = m\,B_{0 \|}^{*}$ (i.e., $d^{3}x\,d^{3}p = \cal{J}_{gc}\,d^{3}R\,dp_{\|}\,d\mu\,d\theta$). The unperturbed guiding-center Hamiltonian dynamics is, thus, expressed in terms of the Hamiltonian (\ref{eq:gc_eH0}) and the Poisson bracket (\ref{eq:gc_ePB}) as $\dot{\cal{Z}}^{\alpha} \equiv \{ \cal{Z}^{\alpha},\; \cal{H}_{gc} \}_{gc}$. In particular, the invariance condition $\dot{\mu} \equiv 0$ for the guiding-center magnetic moment follows from the fact that the guiding-center Hamiltonian 
(\ref{eq:gc_eH0}) is independent of the fast gyroangle $\theta$ (to all orders in $\epsilon_{B}$).

\subsection{Perturbed guiding-center Hamiltonian dynamics}

We now consider how the guiding-center Hamiltonian dynamics is affected by the introduction of low-frequency electromagnetic field fluctuations 
$(\Phi_{1},\bf{A}_{1})$. These fluctuations are assumed to satisfy the low-frequency gyrokinetic ordering \cite{Frieman_Chen,Brizard_89}:
\begin{equation}
\epsilon_{\omega} \;\equiv\; \frac{\omega}{\Omega_{0}} \;\simeq\; \frac{k_{\|}}{|\bf{k}_{\bot}|} \;\ll\; 1 \;\;\;{\rm and}\;\;\;
|\bf{k}_{\bot}|\,\rho \;\simeq\; 1,
\label{eq:gk_ord}
\end{equation}
where $\Omega_{0} \equiv eB_{0}/mc$ denotes the charged-particle's gyrofrequency and $\epsilon_{\omega}$ is a small dimensionless ordering parameter associated with the electromagnetic perturbations space-time scales, with $\omega$ denoting the characteristic wave frequency, $k_{\|}$ the characteristic parallel wavenumber and $\bf{k}_{\bot}$ the characteristic perpendicular wavevector (both with respect to the unperturbed magnetic field $\bf{B}_{0}$).

Under the electromagnetic perturbations $(\Phi_{1},\bf{A}_{1})$, the guiding-center phase-space Lagrangian (\ref{eq:gc_epsl0}) and Hamiltonian 
(\ref{eq:gc_eH0}) become
\begin{equation}
\Gamma_{gc}^{\prime} \;\equiv\; \Gamma_{gc 0} \;+\; \epsilon_{\delta}\,\Gamma_{gc 1} \;\;\; {\rm and}\;\;\; \cal{H}_{gc}^{\prime} \;\equiv\; 
\cal{H}_{gc 0} \;+\; \epsilon_{\delta}\, \cal{H}_{gc 1},
\label{eq:pgc_epsl}
\end{equation}
where the zeroth-order guiding-center phase-space Lagrangian $\Gamma_{gc 0}$ and Hamiltonian $\cal{H}_{gc 0}$ are given by (\ref{eq:gc_epsl0}) and 
(\ref{eq:gc_eH0}), respectively. In what follows, although the three small parameters $(\epsilon_{B},\epsilon_{\delta},\epsilon_{\omega})$ may be of the same order in practice, we keep them separate in order to retain the correct quadratic nonlinearities in the reduced Hamiltonian dynamics. In 
Eq.~(\ref{eq:pgc_epsl}), the first-order guiding-center phase-space Lagrangian $\Gamma_{gc 1}$ and Hamiltonian $\cal{H}_{gc 1}$ are
\begin{eqnarray}
\Gamma_{gc 1} & = & \frac{e}{c}\,\bf{A}_{1}(\bf{R} + \vb{\rho})\bdot d(\bf{R} + \vb{\rho}) \;\equiv\; \frac{e}{c}\,
\bf{A}_{1gc}(\bf{R};\mu,\theta)\bdot d(\bf{R} + \vb{\rho}), \label{eq:gamma1_gc} \\
 &  & \nonumber \\
\cal{H}_{gc 1} & = & e\Phi_{1}(\bf{R} + \vb{\rho}) \;\equiv\; e\Phi_{gc}(\bf{R};\mu,\theta), \label{eq:Ham1_gc}
\end{eqnarray}
where $\bf{A}_{1gc}(\bf{R};\mu,\theta)$ and $\Phi_{1gc}(\bf{R};\mu,\theta)$ denote perturbation potentials evaluated at a particle's position $\bf{x} \equiv \bf{R} + \vb{\rho}$ expressed in terms of the guiding-center position $\bf{R}$ and the gyroangle-dependent gyroradius vector $\vb{\rho}(\mu,
\theta)$. Here, the time dependence of $(\Phi_{1},\bf{A}_{1})$ is not shown explicitly for simplicity and, to lowest order in $\epsilon_{B}$, we ignore the spatial dependence of $\vb{\rho}$. 

Because of the gyroangle-dependence in the guiding-center perturbation potentials $(\Phi_{1gc}, \bf{A}_{1gc})$, the guiding-center magnetic moment 
$\mu$ is no longer conserved by the perturbed guiding-center equations of motion, i.e., $\dot{\mu} = \cal{O}(\epsilon_{\delta})$. To remove the gyroangle-dependence from the perturbed guiding-center phase-space Lagrangian and Hamiltonian (\ref{eq:gamma1_gc})-(\ref{eq:Ham1_gc}), we proceed with the time-dependent {\it gyrocenter} phase-space transformation 
\[ \cal{Z} \;\equiv\; (\bf{R},p_{\|},\mu,\theta,w,t) \;\;\rightarrow\;\; \ov{\cal{Z}} \;\equiv\; (\ov{\bf{R}},\ov{p}_{\|},\ov{\mu},\ov{\theta},
\ov{w},t), \]
where $\ov{\cal{Z}}$ denote the \textit{gyrocenter} (gy) extended phase-space coordinates; we note that $\ov{p}_{\|}$ represents a mixed-canonical momentum coordinate (as will be shown later) and the time coordinate $t$ is not affected by this transformation. 

The results of the nonlinear Hamiltonian gyrocenter perturbation analysis \cite{Brizard_89} are summarized as follows. To first order in the small amplitude parameter $\epsilon \equiv \epsilon_{\delta}$ and zeroth order in the space-time-scale parameters $(\epsilon_{\omega}, \epsilon_{B})$, this transformation is represented in terms of generating vector fields $(\cal{G}_{1},\cal{G}_{2},...)$ as
\begin{equation}
\ov{\cal{Z}}^{\alpha} \;\equiv\; \cal{Z}^{\alpha} \;+\; \epsilon\,\cal{G}_{1}^{\alpha} \;+\; \cdots
\end{equation}
Here, the gyrocenter phase-space Lagrangian is chosen to be of the form 
\begin{equation}
\ov{\Gamma} \;\equiv\; \ov{\Gamma}_{gc} \;=\; \frac{e}{c}\,\bf{A}_{0}^{*}\bdot d\ov{\bf{R}} \;+\; (mc/e)\, \ov{\mu} \,d\ov{\theta} \;-\; 
\ov{w}\; dt,
\label{eq:gamma_gy}
\end{equation}
where $\bf{A}_{0}^{*} \equiv \bf{A}_{0} + (c/e)\,\ov{p}_{\|}\,\bhat_{0}$, so that the gyrocenter Poisson bracket $\{\;,\;\}_{\ov{\cal{Z}}} \equiv 
\{\;,\;\}_{\cal{Z}}$ has the same form as the unperturbed guiding-center Poisson bracket (\ref{eq:gc_ePB}). 

As calculated previously in Eq.~(\ref{eq:G1_def}), the components of the first-order gyrocenter generating vector field $\cal{G}_{1}$ are
\begin{equation}
\cal{G}_{1}^{a} \;\equiv\; \left\{ S_{1},\; \cal{Z}^{a} \right\} \;+\; \frac{e}{c}\,\bf{A}_{1gc}\bdot \left\{ \bf{R} + \vb{\rho},\; \cal{Z}^{a} 
\right\},
\label{eq:G_gy1}
\end{equation}
where the first-order scalar field $\cal{S}_{1}$ is determined as follows. The first-order gyrocenter Hamiltonian is determined from the first-order Lie-transform equation (\ref{eq:eham_1})
\[ \ov{\cal{H}}_{1} \;=\; e \left( \Phi_{1gc} \;-\; \bf{A}_{1gc}\bdot\frac{{\bf v}}{c} \right) \;-\; \{ \cal{S}_{1},\; \cal{H}_{0} \} \;\equiv\;
e\,\psi_{1gc} \;-\; \{ \cal{S}_{1},\; \cal{H}_{0} \}, \]
were $\psi_{1gc} \equiv \Phi_{1 gc} - \bf{A}_{1 gc}\bdot\bf{v}/c$ defines an effective first-order perturbation potential. The gyroangle-averaged part of this first-order equation yields $\cal{H}_{1} \equiv e\,\langle\psi_{1gc}\rangle$, while the solution for the scalar field $\cal{S}_{1}$ is 
\[ \cal{S}_{1} \;=\; \frac{e}{\Omega_{0}}\;\int \wt{\psi}_{1gc}\,d\ov{\theta} \;\equiv\; \frac{e}{\Omega_{0}}\;\wt{\Psi}_{1gc}, \]
where $\wt{\psi}_{1gc} \equiv \psi_{1gc} - \langle\psi_{1gc}\rangle$ denotes the gyroangle-dependent part of the first-order effective potential 
$\psi_{1gc}$. Next, the second-order term in the gyrocenter Hamiltonian is expressed in terms of Eq.~(\ref{eq:H2_average}) as
\[ \ov{\cal{H}}_{2} \;=\; \frac{e^{2}}{2mc^{2}}\; \left\langle|\bf{A}_{1gc}|^{2} \right\rangle \;-\; \frac{e^{2}}{2\Omega_{0}}\; 
\left\langle\left\{ \wt{\Psi}_{1gc},\; \wt{\psi}_{1gc} \right\} \right\rangle, \]
where $\{\bf{R} + \vb{\rho}, {\bf v}\} = {\bf I}/m$ and $\{\bf{R} + \vb{\rho}, \Phi_{1gc}\} = 0 = \{\bf{R} + \vb{\rho}, \bf{A}_{1gc}\}$ were used. 

Up to second order in the amplitude parameter $\epsilon$, the extended phase-space gyrocenter Hamiltonian is, therefore, expressed as
\begin{equation}
\ov{\cal{H}} \;=\; \ov{\cal{H}}_{0} + \epsilon\;e \langle\psi_{1gc}\rangle + \frac{\epsilon^{2}}{2} \left( 
\frac{e^{2}}{mc^{2}}\; \left\langle|\bf{A}_{1gc}|^{2} \right\rangle \;-\; \frac{e^{2}}{\Omega_{0}}\; \left\langle\left\{ \wt{\Psi}_{1gc},\;
\wt{\psi}_{1gc} \right\} \right\rangle \right),
\label{eq:gy_H}
\end{equation} 
where $\ov{\cal{H}}_{0} = \ov{p}_{\|}^{2}/2m + \ov{\mu}\,B_{0} - \ov{w}$ denotes the unperturbed extended guiding-center Hamiltonian and the gyrocenter parallel momentum
\begin{equation} 
\ov{p}_{\|} \;=\; p_{\|} \;+\; \epsilon\; \frac{e}{c}\,\bf{A}_{1gc}\bdot\bhat_{0} \;+\; \cal{O}(\epsilon\epsilon_{\omega}, \epsilon
\epsilon_{B})
\label{eq:gyro_can}
\end{equation}
is a \textit{mixed}-canonical momentum coordinate (i.e., it is kinetic with respect to the background vector potential $\bf{A}_{0}$ and canonical with respect to the perturbed vector potential $\bf{A}_{1}$). Lastly, to first order in $\epsilon$, the low-frequency gyrocenter push-forward operator is defined as
\begin{equation}
{\sf T}_{\epsilon}^{-1}\cal{F} \;=\; \cal{F} \;-\; \epsilon\, \left( \frac{e}{\Omega_{0}} \left\{ \wt{\Psi}_{1gc},\; \cal{F} \right\} \;+\; 
\frac{e}{c}\,\bf{A}_{1gc}\bdot \left\{ \ov{\bf{R}} + \ov{\vb{\rho}},\; \cal{F} \right\} \right) \;+\; \cal{O}(\epsilon^{2}).
\label{eq:push_gy}
\end{equation}
To lowest order in magnetic-field nonuniformity (i.e., up to $\epsilon_{B}^{0}$), the push-forward operator (\ref{eq:push_gy}) is expressed as
\[ {\sf T}_{\epsilon}^{-1}\cal{F} \;=\; \cal{F} \;-\; \left. \left. \frac{e}{B_{0}} \right( \psi_{1gc} \;-\; 
\langle\psi_{1gc}\rangle \right) \pd{\cal{F}}{\ov{\mu}} \;-\; \frac{e}{c}\;\bf{A}_{1gc}\bdot\left( 
\frac{\Omega_{0}}{B_{0}}\;\pd{\ov{\vb{\rho}}}{\ov{\theta}}\;\pd{\cal{F}}{\ov{\mu}} \;+\;
\bhat_{0}\;\pd{\cal{F}}{\ov{p}_{\|}} \right), \]
which clearly exhibits the standard nonadiabatic (first term) and adiabatic (second and third terms) parts \cite{Catto_Tang_Baldwin,Brizard_94} of the guiding-center Vlasov distribution.

We have, thus, obtained a reduced (gyroangle-independent) gyrocenter Hamiltonian description of charged-particle motion in nonuniform magnetized plasmas perturbed by low-frequency electromagnetic fluctuations. At this level, the nonlinear gyrokinetic Vlasov equation can be used to study the evolution of a distribution of \textit{test}-gyrocenters in the presence of low-frequency electromagnetic fluctuations. For a self-consistent treatment that include an electromagnetic field response to the gyrocenter Hamiltonian dynamics, a set of low-frequency Maxwell's equations with charge and current densities expressed in terms of moments of the gyrocenter Vlasov distribution is required.

\section{\label{sec:vp}Variational Principles for Exact and Reduced \\ Vlasov-Maxwell Equations}

In this Section, we plan to derive the nonlinear self-consistent gyrokinetic Vlasov-Maxwell equations from a reduced variational principle. This reduced variational principle will also be used to derive an exact energy conservation law for the reduced Vlasov-Maxwell equations. We begin this Section with the variational principle for the exact Vlasov-Maxwell equations \cite{Brizard_00a}.

\subsection{Exact Vlasov-Maxwell equations}

The variational principle for the Vlasov-Maxwell equations is expressed in terms of the action functional 
\cite{Brizard_00a}
\begin{equation}
\cal{A} \;=\; -\,\int d^{8}\cal{Z}\;\cal{F}(\cal{Z})\; \cal{H}(\cal{Z}) \;+\; \int \frac{d^{3}x\,dt}{16\pi}\; {\sf F}:{\sf F},
\label{eq:action_exact}
\end{equation} 
where $F_{\mu\nu} = \partial_{\mu}A_{\nu} - \partial_{\nu}A_{\mu}$ denotes the electromagnetic four-tensor (${\sf F} = {\sf d}A$ is an exact two-form); here, we introduced the covariant notation $A_{\mu} = (-\,\Phi,\bf{A})$ and $x^{\mu} = (ct,\bf{x})$, with the Minkowski space-like metric $g_{\mu\nu} = 
{\rm diag}(-1,1,1,1)$. We establish the connection between the extended Vlasov distribution $\cal{F}(\cal{Z})$ and the time-dependent Vlasov distribution 
$f(\bf{z},t)$ on the six-dimensional phase space $\bf{z} = (\bf{x},\bf{v})$ by imposing the physical condition $w = H(\bf{z},t)$ on the extended phase-space Vlasov distribution $\cal{F}(\cal{Z})$:
\begin{equation}
\cal{F}(\cal{Z}) \;\equiv\; \delta[w - H(\bf{z},t)]\;f(\bf{z},t).
\label{eq:ext_vlasov}
\end{equation}
Next, the variational principle $\delta\cal{A} = 0$ based on the action functional (\ref{eq:action_exact}) considers 
{\it constrained} Eulerian variations for the extended Vlasov distribution $\cal{F}$ defined as
\begin{equation} 
\delta\cal{F} \;\equiv\; \Delta\cal{F} \;-\; \delta\cal{Z}^{a}\,\partial_{a}\cal{F} \;=\; -\; \delta\cal{Z}^{a}\,
\partial_{a}\cal{F},
\label{eq:Euler_Vlasov}
\end{equation}
where the Lagrangian variation $\Delta\cal{F}$ is identically zero for the extended Vlasov distribution $\cal{F}$ and the extended phase-space virtual displacement $\delta\cal{Z}$, given by Eq.~(\ref{eq:eHam_S}), is generated by the scalar field $\cal{S}$ and variations in vector potential 
$\delta\bf{A}$: 
\begin{equation}
\delta\cal{Z}^{a} \;=\; \left\{ \cal{Z}^{a},\; \cal{S} \right\}_{\cal{Z}} \;-\; \frac{e}{c}\;\delta\bf{A}\bdot\left\{
\bf{x},\; \cal{Z}^{a} \right\}_{\cal{Z}}.
\label{eq:delta_Z}
\end{equation}
From this definition, the Eulerian variation (\ref{eq:Euler_Vlasov}) of the extended Vlasov distribution is
\begin{equation}
\delta\cal{F} \;\equiv\; \{ \cal{S},\; \cal{F}\}_{\cal{Z}} \;+\; \frac{e}{c}\;\delta\bf{A}\bdot\{{\bf x},\; 
\cal{F} \}_{\cal{Z}},
\label{eq:vlas_cons}
\end{equation}
Under Eulerian variations of the electromagnetic potentials $\delta A^{\mu} = (\delta\Phi,\delta\bf{A})$ and the Eulerian variation (\ref{eq:vlas_cons}) of the Vlasov distribution $\cal{F}$, the variation of the action functional (\ref{eq:action_exact}), $\delta\cal{A} \equiv \int \delta\cal{L}\,d^{3}x 
dt$, can be expressed in terms of the variation of the Lagrangian density
\begin{eqnarray}
\delta\cal{L} & = & \pd{}{x^{\nu}} \left[\; \frac{1}{4\pi}\;\delta A_{\mu}\;F^{\mu\nu} \;-\; \int d^{3}v\,dw\; \cal{S}
\left( \frac{\cal{H}}{m}\;\pd{\cal{F}}{v_{\nu}} \right) \;\right] \;-\;\int d^{3}v\,dw\;\cal{S}\;\{\cal{F},\; 
\cal{H}\}_{\cal{Z}} \nonumber \\
 &  &\mbox{}+\; \delta A_{\nu} \left[\; \frac{1}{4\pi}\; \pd{F^{\mu\nu}}{x^{\mu}} \;+\; \frac{e}{c}\,\int d^{3}v\,dw\; \left\{ x^{\nu},\;\cal{H}\right\}_{\cal{Z}}\;\cal{F}\;\right].
\label{eq:Lagvar_exact}
\end{eqnarray}
Stationarity of the action functional (\ref{eq:action_exact}) with respect to arbitrary virtual phase-space displacements generated by $\cal{S}$ yields the Vlasov equation in extended phase space: 
\begin{equation}
\{ \cal{F},\; \cal{H} \}_{\cal{Z}} \;\equiv\; 0.
\label{eq:veq_ext}
\end{equation}
Substituting Eq.~(\ref{eq:ext_vlasov}) into the extended Vlasov equation (\ref{eq:veq_ext}) yields the standard Vlasov equation in $(6+1)$ phase space:
\begin{equation}
\left( \pd{}{t} \;+\; \frac{d\bf{z}}{dt}\bdot\pd{}{\bf{z}}\right)\;f(\bf{z},t) \;=\; 0.
\label{eq:vlas_eq}
\end{equation}
Next, under general variations $\delta A_{\nu}$ of the electromagnetic four-potential, stationarity of the action functional (\ref{eq:action_exact}) yields the Maxwell equations 
\begin{equation}
\pd{F^{\mu\nu}}{x^{\mu}} \;=\; -\;4\pi\,\frac{e}{c}\int d^{3}v\,dw\; \left\{ x^{\nu},\;\cal{H}\right\}_{\cal{Z}}\;\cal{F}
\;=\; -\;4\pi\,\frac{e}{c}\,\int d^{3}v\; v^{\nu}\;f(\bf{x},\bf{v},t),
\label{eq:Maxwell_exact}
\end{equation}
where the $w$-integration, using Eq.~(\ref{eq:ext_vlasov}), yields the standard form of the Maxwell equations, with $v^{\nu} = (c,\,{\bf v})$. The remaining Maxwell's equations $\partial_{\mu}F_{\nu\sigma} + \partial_{\nu}F_{\sigma\mu} + \partial_{\sigma}F_{\mu\nu} = 0$ follow from the fact that 
${\sf F} = {\sf d}{\sf A}$ is an exact two-form (i.e., ${\sf d}{\sf F} = 0$).

\subsection{Noether equation for the exact Vlasov-Maxwell equations}

Since the Vlasov-Maxwell equations (\ref{eq:vlas_eq})-(\ref{eq:Maxwell_exact}) hold for arbitrary phase-space variations generated by $\cal{S}$ and arbitrary four-potential variations $\delta A_{\nu}$, the variation of the Vlasov-Maxwell Lagrangian density (\ref{eq:Lagvar_exact}) is now expressed as a space-time divergence known as the Noether equation:
\begin{equation}
\delta\cal{L} \;=\; \pd{}{x^{\nu}} \left[\; \frac{1}{4\pi}\;\delta A_{\mu}\;F^{\mu\nu} \;-\; \int d^{3}v\,dw\; \cal{S}
\left( \frac{\cal{H}}{m}\;\pd{\cal{F}}{v_{\nu}} \right) \;\right]. 
\label{eq:noether_exact}
\end{equation}
From the Noether equation, we now derive the energy-momentum conservation law 
\begin{equation}
\partial_{\mu}\;{\sf T}^{\mu\nu} \;\equiv\; 0 
\label{eq:T_VM}
\end{equation}
for the Vlasov-Maxwell equations (\ref{eq:vlas_eq})-(\ref{eq:Maxwell_exact}), where $T^{\mu\nu}$ denotes the Vlasov-Maxwell energy-momentum tensor. First, we consider arbitrary space-time translation $x^{\mu} \rightarrow x^{\mu} + \delta x^{\mu}$ generated by the scalar field $\cal{S}$ defined as 
\begin{equation}
\cal{S} \;\equiv [\,m\bf{v} \;+\; (e/c)\,\bf{A}\,]\bdot\delta\bf{x} \;-\; w\;\delta t, 
\label{eq:S_exact}
\end{equation}
and potential variations $\delta\cal{A}_{\mu}$ defined as
\begin{equation}
\delta A_{\mu} \;\equiv\; {\sf F}_{\mu\nu}\;\delta x^{\nu} \;-\; \partial_{\mu}( A_{\nu}\;\delta x^{\nu}).
\label{eq:A_exact}
\end{equation}
Hence, using the spatial components of Eq.~(\ref{eq:A_exact}): 
\begin{equation} 
\delta\bf{A} \;=\; c\delta t\,(\bf{E} + \nabla\Phi) \;+\; \delta\bf{x}\btimes\bf{B} \;-\; \nabla(\bf{A}\bdot\delta\bf{x}),
\label{eq:mag_exact}
\end{equation}
the space-time translation $\delta x^{\mu}$ is expressed as
\[ \delta x^{\mu} \;=\; (c\,\delta t,\; \delta\bf{x}) \;\equiv\; \{ x^{\mu},\; \cal{S}\}_{\cal{Z}} \;-\; \frac{e}{c}\,
\delta\bf{A}\bdot\{ \bf{x},\; x^{\mu}\}_{\cal{Z}}, \]
which follows from Eq.~(\ref{eq:delta_Z}); note that the phase-space virtual displacements generated by Eqs.~(\ref{eq:S_exact}) and (\ref{eq:mag_exact}) do not affect the velocity-energy coordinates (i.e., $\delta\bf{v} = 0 = \delta w$). Substituting Eqs.~(\ref{eq:S_exact})-(\ref{eq:A_exact}) and 
$\delta\cal{L} \equiv -\;\partial_{\mu}(\delta x^{\mu}\;\cal{L})$ into the Noether equation (\ref{eq:noether_exact}), we find 
\[ 0 \;=\; \pd{}{x^{\nu}} \left[ \frac{\delta x_{\mu}}{4\pi} \left( \frac{g^{\mu\nu}}{4}\;{\sf F}:{\sf F} \;-\; 
F^{\mu}_{\;\;\sigma}\;F^{\sigma\nu} \right) \;+\; \int d^{3}v\;\left( m\bf{v}\bdot\delta\bf{x} \;-\; 
\frac{m}{2}\,|\bf{v}|^{2}\;\delta t \right) v^{\nu}f \right], \]
after the $w$-integration has been carried out and the terms
\[ \pd{}{x^{\nu}} \left[\; \frac{A_{\sigma}\,\delta x^{\sigma}}{4\pi} \left( \pd{F^{\mu\nu}}{x^{\mu}} \;+\; 4\pi\,\frac{e}{c}\,\int d^{3}v\; v^{\nu}\;
f(\bf{x},\bf{v},t) \right) \;\right] \]
vanish identically as a result of Maxwell's equations (\ref{eq:Maxwell_exact}).

The energy conservation law $(\delta x^{0} = c\,\delta t)$ for the Vlasov-Maxwell equations can, thus, be expressed as
\begin{equation}
\pd{\cal{E}}{t} \;+\; \nabla\bdot\bf{S} \;=\; 0,
\label{eq:energy_exact}
\end{equation}
where the energy density $\cal{E}$ and energy-density flux $\bf{S}$ are
\begin{eqnarray}
\cal{E}(\bf{x},t) & = & \frac{1}{8\pi} \left( |\bf{E}|^{2} \;+\; |\bf{B}|^{2} \right) \;+\; 
\int d^{3}v\;\frac{m}{2}\,|\bf{v}|^{2}\;f(\bf{x},\bf{v},t), \label{eq:density_exact} \\
\bf{S}(\bf{x},t) & = & \frac{c}{4\pi}\;\bf{E}\btimes\bf{B} \;+\; \int d^{3}v\;\frac{m}{2}\,|\bf{v}|^{2}\,\bf{v}\;f(\bf{x},\bf{v},t). 
\label{eq:flux_exact}
\end{eqnarray}
The momentum conservation law for the Vlasov-Maxwell equations, on the other hand, can be expressed as
\begin{equation}
\pd{\bf{\Pi}}{t} \;+\; \nabla\bdot{\sf T} \;=\; 0,
\label{eq:momentum_exact}
\end{equation}
where the momentum density $\bf{\Pi}$ and momentum-stress tensor ${\sf T}$ are
\begin{eqnarray}
\bf{\Pi}(\bf{x},t) & = & \frac{\bf{E}\btimes\bf{B}}{4\pi\,c} \;+\; 
\int d^{3}v\;m{\bf v}\;f(\bf{x},\bf{v},t), \label{eq:momdens_exact} \\
{\sf T}(\bf{x},t) & = & \frac{\bf{I}}{8\pi} \left( |\bf{E}|^{2} \;+\; |\bf{B}|^{2} \right) \;-\; \frac{1}{4\pi}
\left( \bf{E}\,\bf{E} \;+\; \bf{B}\,\bf{B} \right) \nonumber \\
 &  &\mbox{}+\; \int d^{3}v\;m\,\bf{v}\bf{v}\;f(\bf{x},\bf{v},t). 
\label{eq:stress_exact}
\end{eqnarray}
The exact Vlasov-Maxwell equations (\ref{eq:vlas_eq})-(\ref{eq:Maxwell_exact}) have, thus, been derived by a variational principle based on the action functional (\ref{eq:action_exact}) involving the extended Vlasov distribution $\cal{F}$ and the four-potential $A_{\mu}$. Exact energy-momentum conservation laws have also been derived by an application of the Noether method.

The covariance of the Vlasov part of the action functional (\ref{eq:action_exact}) is used next to construct the gyrokinetic Vlasov action functional in which the extended particle phase-space coordinates $\cal{Z}$, the extended Vlasov distribution $\cal{F}$, and the extended particle Hamiltonian 
$\cal{H}$ are replaced with their gyrocenter equivalents.

\subsection{Nonlinear low-frequency gyrokinetic Vlasov-Maxwell equations}

The reduced action functional for the low-frequency gyrokinetic Vlasov-Maxwell equations \cite{Brizard_00b,Sugama} is
\begin{equation}
\cal{A}_{R} \;=\; -\;\int d^{8}\cal{Z}\;\cal{F}(\cal{Z})\;\cal{H}(\cal{Z}) \;+\; \int\frac{d^{4}x}{8\pi}\;\left( 
|\nabla\Phi|^{2} \;-\; |\bf{B}|^{2} \right),
\label{eq:action_ngk}
\end{equation}
where we, henceforth, use the notation
\[ \Phi \;\equiv\; \epsilon\,\Phi_{1} \;\;\;{\rm and}\;\;\; \bf{B} \;\equiv\; \bf{B}_{0} \;+\; \epsilon\,\nabla\btimes\bf{A}_{1}, \]
and we omit the overbar to denote gyrocenter coordinates and functions on extended gyrocenter phase space. The absence of the inductive part $-\,c^{-1}\partial_{t}\bf{A}_{1}$ of the perturbed electric field $\bf{E}_{1}$ in the Maxwell part of the reduced action functional (\ref{eq:action_ngk}) means that the inductive current $\partial_{t}\bf{E}_{1}$ will be absent from Amp\`{e}re's equation; this is consistent with the low-frequency approximation 
$(\epsilon_{\omega} \ll 1)$ used in nonlinear gyrokinetic ordering (\ref{eq:gk_ord}).

The variational principle $\delta\cal{A}_{R} = \int \delta\cal{L}_{R}\,d^{4}x \equiv 0$ for the nonlinear low-frequency gyrokinetic Vlasov-Maxwell equations is based on Eulerian variations for $\cal{F}(\cal{Z})$ while variations of the electromagnetic potentials $(\Phi,\bf{A})$ are {\it restricted} to variations of the perturbation potentials $\Phi_{1}(\bf{x},t)$ and $\bf{A}_{1}(\bf{x},t)$ only. Variation of $\cal{A}_{R}$ with respect to 
$\delta\cal{F}(\cal{Z})$ and $\delta A^{\mu}(\bf{x}) = (\delta\Phi_{1},\;\delta\bf{A}_{1})$ yields
\begin{eqnarray}
\delta\cal{A}_{R} & = & -\;\int d^{8}\cal{Z}\left[\;\delta\cal{F}(\cal{Z})\; \cal{H} \;+\; \cal{F}(\cal{Z})\, 
\int d^{3}x \left( \delta A_{1\mu}(\bf{x})\;\fd{H}{A_{1\mu}(\bf{x})} \right) \;\right] \nonumber \\
 &   &\mbox{}+\; \left. \left. \int \frac{d^{4}x}{4\pi} \right(\; \epsilon\; \nabla\delta \Phi_{1}\bdot
\nabla\Phi \;-\; \epsilon\, \nabla\btimes
\delta\bf{A}_{1}\bdot\bf{B} \;\right).
\label{eq:var_A}
\end{eqnarray}
Here, the Eulerian variation $\delta\cal{F}$ is constrained to be of the form
\begin{equation}
\delta\cal{F} \;\equiv\; \{ \cal{S},\; \cal{F} \},
\label{eq:dF_can}
\end{equation}
where $\cal{S}$ generates the virtual extended phase-space displacement $\delta\cal{Z}^{a} \equiv \{ \cal{Z}^{a},\; \cal{S}\}$ and $\{\;,\;\}$ is the unperturbed guiding-center Poisson bracket (\ref{eq:gc_ePB}) on extended gyrocenter phase space; note that since the gyrocenter momentum coordinates are canonical with respect to the magnetic perturbation $\bf{A}_{1}$, the Eulerian variation (\ref{eq:dF_can}) has the canonical form. The functional derivatives 
$\delta H/\delta A_{1\mu}(\bf{x})$ in Eq.~(\ref{eq:var_A}), on the other hand, are evaluated using the gyrocenter Hamiltonian (\ref{eq:gy_H}) (to second order in $\epsilon$) as
\begin{equation}
\fd{H}{A_{1\mu}(\bf{x})} \;\equiv\; -\;\epsilon\; e\left\langle {\sf T}_{\epsilon}^{-1} \left( \frac{v^{\mu}}{c}\,
\delta^{3}_{gc} \right) \right\rangle,
\label{eq:funcdev_A}
\end{equation}
where $\delta^{3}_{gc} \equiv \delta^{3}(\bf{x} - \bf{R} - \vb{\rho})$, the push-forward operator ${\sf T}_{\epsilon}^{-1}$ is given to first order by 
Eq.~(\ref{eq:push_gy}), and we used the identity
\[ A_{1\mu}(\bf{R} + \vb{\rho}) \;=\; \int d^{3}x\;\delta^{3}(\bf{x} - \bf{R} - \vb{\rho})\;A_{1\mu}(\bf{x})
\;\;\;\rightarrow\;\;\; \fd{A_{1gc}^{\mu}}{A_{1\nu}({\bf x})} \;=\; \delta^{\mu\nu}\;\delta^{3}_{gc}. \]
After re-arranging and integrating by parts, the variation (\ref{eq:var_A}) becomes
\begin{eqnarray}
\delta\cal{A}_{R} & = & -\,\int d^{4}x\;\epsilon\,\frac{\delta\Phi_{1}(\bf{x})}{4\pi} \left[\; \nabla^{2}\Phi 
\;+\; 4\pi e\;\int d^{6}Z\; f\;\left\langle {\sf T}_{\epsilon}^{-1}\delta^{3}_{gc}\right\rangle\;\right]
\nonumber \\
 & + & \int d^{4}x\; \epsilon\,\frac{\delta\bf{A}_{1}(\bf{x})}{4\pi} \bdot \left[\; \nabla\btimes\bf{B} \;-\; 4\pi e\; \int d^{6}Z\; f\; 
\left\langle {\sf T}_{\epsilon}^{-1} \left( \frac{\bf{v}}{c}\, \delta^{3}_{gc} \right) \right\rangle \;\right] \nonumber \\
 & + & \int d^{8}\cal{Z}\; \cal{S}\;\{ \cal{F},\; (w - H)\}_{\cal{Z}} \;+\; \int d^{4}x\; \left(\partial\cdot\cal{J}\right),
\label{eq:del_A}
\end{eqnarray}
where we have used $\cal{F}(\cal{Z}) \equiv \delta(w - H)\,f(\bf{Z},t)$ for the gyrocenter Vlasov distribution function in extended phase space in the first two terms in $\delta\cal{A}$, while the last term in Eq.~(\ref{eq:del_A}) involves the exact space-time divergence
\begin{equation}
\partial\cdot\cal{J}(x) \;\equiv\; \pd{}{x^{\mu}}\left(\int d^{8}\cal{Z}\,\delta^{4}(x - R) \;\cal{S}\;\cal{F}\;\dot{R}^{\mu} \right) \;+\; \nabla\bdot \left(\; \epsilon\, \frac{\delta\Phi_{1}}{4\pi}\; \nabla\Phi \;-\; \epsilon\, \frac{\delta\bf{A}_{1}}{4\pi}\btimes\bf{B} \;\right),
\label{eq:var_L}
\end{equation}
where $\dot{R}^{\mu} \equiv \{ R^{\mu},\; \cal{H} \}_{\cal{Z}}$ denotes the lowest-order gyrocenter four-velocity. Since Eq.~(\ref{eq:var_L}) is an exact space-time divergence, it does not contribute to the reduced variational principle $\delta\cal{A}_{R} \equiv 0$. 

By requiring that the action functional $\cal{A}_{R}$ be stationary with respect to arbitrary variations $\cal{S}$ and $\delta A_{1}^{\mu}$ (which vanish on the integration boundaries), we find the nonlinear gyrokinetic Vlasov equation 
\begin{equation}
0 \;=\; \{ \cal{F},\; \cal{H} \},
\label{eq:gk_vlasov}
\end{equation}
and the gyrokinetic Maxwell equations: the gyrokinetic Poisson equation
\begin{equation}
\nabla^{2}\Phi(\bf{x}) \;=\; -\,4\pi\,e\;\int d^{6}Z\; f \;\left\langle {\sf T}_{\epsilon}^{-1}\;
\delta^{3}_{gc} \right\rangle \;\equiv\; -\,4\pi\,e\;\int d^{6}Z\; \left\langle \delta^{3}_{gc}\; {\sf T}_{\epsilon}f \right\rangle,
\label{eq:gk_poisson}
\end{equation}
and the gyrokinetic Amp\`{e}re equation
\begin{equation}
\nabla\btimes\bf{B}(\bf{x}) \;=\; \frac{4\pi\,e}{c}\;\int d^{6}Z\; f(Z)\; \left\langle {\sf T}_{\epsilon}^{-1} \left( 
\bf{v}\,\delta^{3}_{gc} \right) \right\rangle \;\equiv\; \frac{4\pi\,e}{c}\;\int d^{6}Z\; \left\langle \bf{v}\,\delta^{3}_{gc}\; {\sf T}_{\epsilon}f \right\rangle.
\label{eq:gk_ampere}
\end{equation}
If we now substitute $\cal{F}(\cal{Z}) \equiv \delta(w - H)\,f(Z,t)$ into $\{ \cal{F}, \;\cal{H} \}_{\cal{Z}} = 0$, we obtain the standard nonlinear gyrokinetic Vlasov equation written explicitly as 
\begin{equation}
\pd{f}{t} \;+\; \left( \frac{\bf{B}_{0}^{*}}{B_{0 \|}^{*}}\,\pd{H}{p_{\|}} \;+\; \frac{c\wh{\bf{b}}_{0}}{eB_{0 \|}^{*}}\btimes \nabla H \right)\bdot\nabla f \;-\; \left( \frac{\bf{B}_{0}^{*}}{B_{0 \|}^{*}}\bdot\nabla H \right)\,
\pd{f}{p_{\|}} \;=\; 0.
\label{eq:gkv_exp}
\end{equation}
The nonlinear equations (\ref{eq:gk_poisson}), (\ref{eq:gk_ampere}), and (\ref{eq:gkv_exp}), with the gyrocenter Hamiltonian (\ref{eq:gy_H}), are the self-consistent nonlinear gyrokinetic Vlasov-Maxwell equations in general magnetic field geometry \cite{Brizard_89}.

\subsection{Gyrokinetic energy conservation law}

We now apply the Noether method on the gyrokinetic action functional (\ref{eq:action_ngk}) to derive an exact gyrokinetic energy conservation law. By substituting Eqs.~(\ref{eq:gk_vlasov}), (\ref{eq:gk_poisson}), and (\ref{eq:gk_ampere}) into Eq.~(\ref{eq:del_A}), the variational equation
$\delta\cal{A} \equiv \int \delta\cal{L}\;d^{4}x$ (we henceforth ignore the subscript $R$) yields the Noether equation
\begin{equation}
\delta\cal{L}(x) \;\equiv\; \partial\cdot\cal{J}(x).
\label{eq:noether}
\end{equation}
In the Noether method, the variations $(\cal{S},\delta A_{1}^{\mu},\delta\cal{L})$ are expressed in terms of generators for infinitesimal translations in space or time.

Following a translation in time $t \rightarrow t + \delta t$, the variations $\cal{S}$, $\delta\Phi_{1}$, $\delta\bf{A}_{1}$, and 
$\delta\cal{L}$ become, respectively, 
\begin{equation}
\left. \begin{array}{rcl}
\cal{S} & = & -\,w\;\delta t \\
 &   & \\
\delta\Phi_{1} & = & -\delta t\,\partial_{t}\Phi_{1} \\
 &   & \\
\delta\bf{A}_{1} & = & -\delta t\,\partial_{t} \bf{A}_{1} \;\equiv\; c\delta t\,(\bf{E} + \nabla\Phi) \\
 &   & \\
\delta\cal{L} & = & -\delta t\,\partial_{t}\cal{L} 
\end{array} \right\}.
\label{eq:time}
\end{equation}
In Eq.~(\ref{eq:time}), the expression for $\cal{S}$ satisfies $\delta t \equiv \{ t,\; \cal{S}\}_{\cal{Z}}$ and the Vlasov-Maxwell Lagrangian density is
\[ \cal{L} \;=\; \frac{1}{8\pi} \left( |\nabla\Phi|^{2} \;-\; |\bf{B}|^{2} \right) \;\equiv\; \frac{1}{4\pi}\,|\nabla\Phi|^{2} \;-\; 
\cal{E}_{EM}, \]
after the physical constraint $\cal{H} = 0$ is imposed in the space-time integrand of the reduced action functional 
(\ref{eq:action_ngk}); here, $\cal{E}_{EM} \equiv (|\nabla\Phi|^{2} + |\bf{B}|^{2})/8\pi$ denotes the electromagnetic-field energy density (in the low-frequency limit). 

By combining Eq.~(\ref{eq:time}) with Eqs.~(\ref{eq:var_L}) and (\ref{eq:noether}), we obtain
\begin{eqnarray}
0 & = & \pd{}{t} \left( \frac{1}{4\pi}\;\overbrace{|\nabla\Phi|^{2}}^{(I)} \;-\; \cal{E}_{EM} \;-\; \int d^{6}Z\; 
\delta^{3}(\bf{x} - \bf{R})\; H\,f \right) \nonumber \\
 &   &\mbox{}-\; \nabla\bdot\left[\; \int d^{6}Z\;\delta^{3}(\bf{x} - \bf{R})\; H\,f\;\dot{\bf{R}} \;\right] \nonumber \\
 &   &\mbox{}+\; \nabla\bdot \left[\; -\, 
\frac{\epsilon}{4\pi}\, \overbrace{\partial_{t}\Phi_{1}\; \nabla\Phi}^{(II)} \;-\; \frac{c}{4\pi}\;( \bf{E} + \overbrace{\nabla\Phi}^{(III)}) 
\btimes\bf{B} \;\right]. 
\end{eqnarray}
The term $(I)$ can be written as
\begin{equation}
\pd{}{t} \left( \frac{1}{4\pi}\;|\nabla\Phi|^{2} \right) \;=\; \pd{}{t} \left[\; \nabla\bdot \left( \frac{\Phi\;\nabla\Phi}{4\pi} \right) \;+\; 
\int d^{6}Z\, \delta^{3}(\bf{x} - \bf{R}) \;f\; \left\langle e\,{\sf T}_{\epsilon}^{-1}\Phi_{gc} \right\rangle \;\right],
\label{eq:I}
\end{equation}
where $\Phi_{gc} = \epsilon\,\Phi_{1gc}$, the identity $|\nabla\Phi|^{2} \equiv \nabla\bdot(\Phi\;\nabla\Phi) - \Phi\; \nabla^{2}\Phi$ was used, and 
Eq.~(\ref{eq:gk_poisson}) was substituted. The term $(II)$ can be written as
\begin{equation}
-\,\nabla\bdot \left(\; \frac{\epsilon}{4\pi}\,\pd{\Phi_{1}}{t}\;\nabla\Phi \;\right) \;=\; -\; \nabla\pd{}{t} \bdot 
\left(\; \frac{\epsilon}{4\pi}\; \Phi_{1}\;\nabla\Phi \;\right) \;+\; \nabla\bdot \left( \frac{\epsilon^{2}}{4\pi}\; 
\Phi_{1}\;\nabla\pd{\Phi_{1}}{t} \right).
\label{eq:II}
\end{equation}
Here, the first term on the right side of Eq.~(\ref{eq:II}) cancels the first term in Eq.~(\ref{eq:I}). Lastly, the term $(III)$ can be written as
\begin{eqnarray}
-\;\nabla\bdot\left(\; \nabla\Phi\,\btimes \frac{c}{4\pi}\; \bf{B} \;\right) & = & \nabla\bdot \left(\; \Phi\; 
\frac{c}{4\pi}\;\nabla \btimes\bf{B} \;\right) \nonumber \\
 & = & \nabla\bdot \left[\; e\; \int d^{6}Z\; f\; \left\langle{\sf T}_{\epsilon}^{-1}\left( \bf{v}\,\Phi_{gc}\right) \right\rangle \;\right],
\label{eq:III}
\end{eqnarray}
where Eq.~(\ref{eq:gk_ampere}) was substituted. By collecting the remaining terms, we obtain the following expression for the {\it local} gyrokinetic energy conservation law:
\begin{equation}
\pd{\cal{E}}{t} \;+\; \nabla\bdot\bf{S} \;=\; 0,
\label{eq:lge_law}
\end{equation}
where the gyrokinetic energy density is
\begin{eqnarray}
\cal{E}(\bf{x},t) & = & \int d^{6}Z\,\delta^{3}(\bf{x} - \bf{R})\; f(Z,t)\, 
\left( H \;-\; e \left\langle {\sf T}_{\epsilon}^{-1}\Phi_{gc} \right\rangle \right) 
\nonumber \\
 &   &\mbox{}+\; \frac{1}{8\pi} \left( |\nabla\Phi|^{2} \;+\; |\bf{B}|^{2} \right), 
\label{eq:E_density}
\end{eqnarray}
while the gyrokinetic energy density flux is
\begin{eqnarray}
\bf{S}(\bf{x},t) & = & \int d^{6}Z\,\delta^{3}(\bf{x} - \bf{R})\; f(Z,t)
\left( H\,\dot{\bf{R}} \;-\; e\left\langle {\sf T}_{\epsilon}^{-1}\;\bf{v}\Phi_{gc} 
\right\rangle \right) \nonumber \\
 &   &\mbox{}+\; \frac{1}{4\pi} \left(\; c\bf{E}\btimes\bf{B} \;-\; 
\epsilon\;\Phi\;\nabla\pd{\Phi_{1}}{t} \;\right). 
\label{eq:E_flux}
\end{eqnarray} 
We obtain the following expression for the \textit{global} gyrokinetic energy conservation law $dE/dt = 0$, where the global gyrokinetic energy is
\begin{equation}
E \;=\; \int \frac{d^{3}x}{8\pi} \left( |\nabla\Phi|^{2} \;+\; |\bf{B}|^{2} \right) \;+\; \int d^{6}Z\,f\, \left( H 
\;-\; e \left\langle {\sf T}_{\epsilon}^{-1}\Phi_{gc} \right\rangle \right).
\label{eq:e_global}
\end{equation}
The existence of this exact global energy conservation law allowed the development of energy-conserving simulation techniques for nonlinear electrostatic \cite{Dubin_etal} and electromagnetic \cite{Hahm_Lee_Brizard,Brizard_89} gyrokinetic equations.

\section{\label{sec:sum}Summary}

Through the use of Lie-transform perturbation methods on extended particle phase space, we have derived a set of nonlinear low-frequency gyrokinetic Vlasov-Maxwell equations describing the reduced Hamiltonian description of gyrocenter dynamics in a time-independent background magnetic field perturbed by low-frequency electromagnetic fluctuations. A self-consistent treatment is obtained through a low-frequency gyrokinetic variational principle and an exact gyrokinetic energy conservation law is obtained by applying the Noether method. Throughout the work, the Lagrangian and Hamiltonian methods (and differential geometry) in eight-dimensional extended phase space were emphasized.

Further developments in gyrokinetic theory not presented here include nonlinear relativistic gyrokinetic Vlasov-Maxwell equations \cite{Brizard_Chan} and nonlinear bounce-center Hamiltonian dynamics \cite{Fong_Hahm,Brizard_00c} (based on the asymptotic elimination of the fast bounce-motion time scale from the gyrocenter Hamiltonian dynamics).

\vspace*{0.2in}

\noindent
\textbf{Acknowledgments}

\vspace*{0.2in}

I wish to thank Pierre-Louis Sulem (Observatoire de Nice) for inviting me to present some of this work at the Workshop on Kinetic Theory held at the Fields Institute from March 29 to April 2, 2004. I also wish to thank Maurizio Ottaviani (CEA Cadarache) for inviting me to Cadarache to give lectures on the foundations of gyrokinetic theory based on material presented here. Lastly, I wish to express my sincere gratitude to Allan Kaufman for his constant support over the past 15 years. The work presented here was supported in part by the U.S. Department of Energy under Contract 
No.~DE-AC03-76SFOO098.


\begin{thebibliography}{99}

\bibitem{Northrop} T.G.~Northrop, \textit{Adiabatic Motion of Charged Particles} (Wiley, New York, 1963).

\bibitem{RGL_83} R.G.~Littlejohn, \textit{Variational principles of guiding centre motion}, J.~Plasma Phys. \textbf{29}, 111-125 (1983).

\bibitem{Taylor} J.B.~Taylor, \textit{Magnetic moment under short-wave electrostatic perturbations}, Phys.~Fluids 
\textbf{10}, 1357-1359 (1967).

\bibitem{Rutherford_Frieman} P.H.~Rutherford and E.A.~Frieman, \textit{Drift instabilities in general magnetic field configurations}, Phys.~Fluids \textbf{11}, 569-585 (1968).

\bibitem{Catto} P.J.~Catto, \textit{Linearized gyrokinetics}, Plasma Phys.~\textbf{20}, 719-722 (1978).

\bibitem{Antonsen_Lane} T.M.~Antonsen and B.~Lane, \textit{Kinetic equations for low frequency instabilities in inhomogeneous plasmas}, Phys.~Fluids \textbf{23}, 1205-1214 (1980).

\bibitem{Catto_Tang_Baldwin} P.J.~Catto, W.M.~Tang, and D.E.~Baldwin, \textit{Generalized gyrokinetics}, Plasma Phys.~\textbf{23}, 639-650 (1981).

\bibitem{Krommes} J.A.~Krommes, \textit{Dielectric response and thermal fluctuations in gyrokinetic plasma}, Phys.~Fluids
\textbf{5}, 1066-1100 (1993).

\bibitem{Brizard_95} A.J.~Brizard, \textit{Nonlinear gyrokinetic Vlasov equation for toroidally rotating axisymmetric tokamaks}, Phys.~Plasmas 
\textbf{2}, 459-471 (1995).

\bibitem{Qin_etal} H.~Qin, W.M.~Tang, G.~Rewoldt, and W.W.~Lee, \textit{On the gyrokinetic equilibrium}, Phys.~Plasmas \textit{7}, 991-1000 (2000).

\bibitem{Frieman_Chen} E.A.~Frieman and L.~Chen, \textit{Nonlinear gyrokinetic equations for low-frequency electromagnetic waves in general plasma equilibria}, Phys.~Fluids \textbf{25}, 502-508 (1982).

\bibitem{Dubin_etal} D.H.E.~Dubin, J.A.~Krommes, C.~Oberman, and W.W.~Lee, \textit{Nonlinear gyrokinetic equations}, 
Phys.~Fluids \textbf{26}, 3524-3535 (1983).

\bibitem{Hahm_Lee_Brizard} T.S.~Hahm, W.W.~Lee, and A.J.~Brizard, \textit{Nonlinear gyrokinetic theory for finite-beta plasmas}, Phys.~Fluids 
\textbf{31}, 1940-1948 (1988).

\bibitem{Hahm_88} T.S.~Hahm, \textit{Nonlinear gyrokinetic equations for tokamak microturbulence}, Phys.~Fluids 
\textbf{31}, 2670-2673 (1988).

\bibitem{Brizard_89} A.J.~Brizard, \textit{Nonlinear gyrokinetic Maxwell-Vlasov equations using magnetic co-ordinates}, 
J.~Plasma Phys. \textbf{41}, 541-559 (1989).

\bibitem{Hahm_96} T.S.~Hahm, \textit{Nonlinear gyrokinetic equations for turbulence in core transport barriers}, 
Phys.~Plasmas \textbf{3}, 4658-4664 (1996).

\bibitem{Lee} W.W.~Lee, \textit{Gyrokinetic particle simulation model}, J.~Comput.~Phys.~\textbf{72}, 243-269 (1987).

\bibitem{Lee_Tang} W.W.~Lee and W.M.~Tang, \textit{Gyrokinetic particle simulation of ion temperature gradient instabilities}, Phys.~Fluids \textbf{31}, 612-624 (1988).

\bibitem{Dimits_etal} A.M.~Dimits, G.~Bateman, M.A.~Bateman, \textit{et al.}, \textit{Comparisons and physics basis of tokamak transport models and turbulence simulations}, Phys.~Plasmas \textbf{7}, 969-983 (2000).

\bibitem{Parker_etal} S.E.~Parker, Y.~Chen, W.~Wan, B.I.~Cohen, and W.M.~Nevins, \textit{Electromagnetic gyrokinetic simulations}, Phys.~Plasmas \textbf{11}, 2594-2599 (2004).

\bibitem{Krommes_report} J.A.~Krommes, \textit{Fundamental statistical descriptions of plasma turbulence in magnetic fields}, Phys.~Rep.~\textbf{360}, 1-352 (2002).

\bibitem{Arnold} V.I.~Arnold, \textit{Mathematical Methods of Classical Mechanics, 2nd ed.} (Springer-Verlag, 1989).

\bibitem{Goldstein} H.~Goldstein, C.~Poole, and J.~Safko, \textit{Classical Mechanics}, 3rd ed. (Addison-Wesley, San Francisco, 2002), Sec.~9.5.

\bibitem{Brizard_01} A.J.~Brizard, \textit{A geometric view of Hamiltonian perturbation theory}, Phys.~Lett.~A 
\textbf{291}, 146-149 (2001).

\bibitem{Spivak} M.~Spivak, \textit{Calculus on Manifolds: A Modern Approach to Classical Theorems of Advanced Calculus} (Westview Press, 1965).

\bibitem{osc} J.R.~Cary and A.N.~Kaufman, \textit{Ponderomotive effects in collisionless plasma: A Lie-transform 
approach}, Phys.~Fluids \textbf{24}, 1238-1250 (1981).

\bibitem{Cary_RGL} J.R.~Cary and R.G.~Littlejohn, \textit{Noncanonical Hamiltonian mechanics and its application to magnetic field line flow}, Ann.~Phys.~(N.Y.) \textbf{151}, 1-34 (1983).

\bibitem{RGL_82} R.G.~Littlejohn, \textit{Hamiltonian perturbation theory in noncanonical coordinates}, J.~Math.~Phys. \textbf{23}, 742-747 (1982).

\bibitem{Abraham_Marsden} R.~Abraham and J.E.~Marsden, \textit{Foundations of Mechanics}, 2nd ed. (Benjamin/Cummings, Reading, MA, 1978). 

\bibitem{Brizard_94} A.J.~Brizard, \textit{Eulerian action principles for linearized reduced dynamical equations}, 
Phys.~Plasmas \textbf{1}, 2460-2472 (1994).

\bibitem{Brizard_00a} A.J.~Brizard, \textit{New variational principle for the Vlasov-Maxwell equations}, 
Phys.~Rev.~Lett.~\textbf{84}, 5768-5771 (2000).

\bibitem{Brizard_00b} A.J.~Brizard, \textit{Variational principle for the nonlinear gyrokinetic Vlasov-Maxwell equations}, 
Phys.~Plasmas.~\textbf{7}, 4816-4822 (2000).

\bibitem{Sugama} H.~Sugama, \textit{Gyrokinetic field theory}, Phys.~Plasmas \textbf{7}, 466-480 (2000).

\bibitem{Brizard_Chan} A.J.~Brizard and A.A.~Chan, \textit{Relativistic nonlinear gyrokinetic Maxwell-Vlasov equations}, Phys.~Plasmas \textbf{6}, 4548-4558 (1999).

\bibitem{Fong_Hahm} B.H.~Fong and T.S.~Hahm, \textit{Bounce-averaged kinetic equations and neoclassical polarization density}, Phys.~Plasmas \textbf{6}, 188-199 (1999).

\bibitem{Brizard_00c} A.J.~Brizard, \textit{Nonlinear bounce-center Hamiltonian dynamics in general magnetic geometry}, Phys.~Plasmas \textbf{7},
3238-3246 (2000).

\end{thebibliography}
\end{document}